\documentclass[]{spie}  

\usepackage[utf8]{inputenc}
\usepackage{amsmath,amsfonts,amssymb}
\usepackage{graphicx}
\graphicspath{{./}}
\usepackage[colorlinks=true, allcolors=blue]{hyperref}
\usepackage{siunitx}
\usepackage{multirow}

\DeclareSIUnit{\mas}{mas}
\newcommand{\etamag}{\ensuremath{\|\eta\|}}

\title{Visible-Light High-Contrast Polarimetry with MagAO-X: Characterization and Initial Results}

\author[a,b]{Miles Lucas}
\affil[a]{Steward Observatory, University of Arizona, Tucson, AZ 87521, USA}
\affil[b]{Subaru Telescope, National Astronomical Observatory of Japan, 650 N. Aohoku Pl., Hilo, HI 96720, USA}

\author[a]{Laird Close}

\author[a]{Jared R. Males}

\author[c]{Tiago Gualberto Bezerra de Souza}
\affil[c]{Nuclear and Energy Research Institute, IPEN-CNEN, S\~ao Paulo, SP, 05508-000, Brazil}

\author[d]{Rodrigo Pereira}
\affil[d]{Institute Prof. Jo\~ao Steiner of Instrumentation and Data Science, S\~ao Paulo, SP, 04043-200, Brazil}

\author[a]{Jialin Li}

\author[e]{Joseph D. Long}
\affil[e]{Center for Computational Astrophysics, Flatiron Institute, 162 5th Avenue, New York, NY 10010, USA}

\author[f]{Jaren N. Ashcraft}
\affil[f]{Department of Physics, University of California, Santa Barbara, CA, 93106, USA}

\author[a]{Kyle Van Gorkom}

\author[a,g,b,h]{Olivier Guyon}
\affil[g]{Wyant College of Optical Sciences, University of Arizona, Tucson, AZ 87521, USA}
\affil[h]{Astrobiology Center, 2 Chome-21-1, Osawa, Mitaka, Tokyo, 181-8588, Japan}

\author[i,a]{Sebastiaan Y. Haffert}
\affil[i]{Leiden Observatory, Leiden University, PO Box 9513, 2300 RA Leiden, The Netherlands}

\author[j]{Alexander D. Hedglen}
\affil[j]{Northrop Grumman Corporation, 600 South Hicks Road, Rolling Meadows, IL 60008, USA}

\author[k]{Rob G. van Holstein}
\affil[k]{European Southern Observatory, Alonso de C\'ordova 3107, Casilla 19001, Vitacura, Santiago, Chile}

\author[a]{Parker T. Johnson}

\author[a,g]{Maggie Kautz}

\author[g]{Jay Kueny}

\author[f]{Briley L. Lewis}

\author[g]{Joshua Liberman}

\author[g]{Jennifer Lumbres}

\author[g]{Eden McEwen}

\author[l]{Avalon L. McLeod}
\affil[l]{Draper Laboratory, 555 Technology Square, Cambridge, MA 02139, USA}
\author[f]{Maxwell A. Millar-Blanchaer}

\author[m]{Lauren Schatz}
\affil[m]{Starfire Optical Range, Kirtland Air Force Base, Albuquerque, NM 87123, USA}

\author[g]{Katie Twitchell}

\author[f]{Manxuan Zhang}

\authorinfo{Further author information: send correspondence to M.L.\\M.L.: E-mail: mileslucas@arizona.edu}

\begin{document}
\maketitle

\begin{abstract}
MagAO-X is a visible-light extreme adaptive optics instrument on the 6.5 meter Magellan Clay Telescope, recently upgraded to enable high-contrast polarimetric differential imaging (PDI) in r', i', and z' filters. Polarimetry is a powerful technique for suppressing unpolarized starlight and isolating the faint, polarized signal scattered by circumstellar dust, but it demands precise calibration of instrumental polarization effects introduced by the telescope and instrument optics. We present an overview of the MagAO-X polarimeter and characterize its polarimetric response using a purpose-built polarization generator that injects light of a known polarization state. From these measurements, we fit a Mueller-matrix model of the instrument and quantify its polarimetric efficiency and instrumental polarization as a function of the k-mirror image rotator angle and observing filter. The initial characterization revealed significant, dynamic inefficiencies driven by the image rotator, motivating the deployment of a dual rotating quarter-wave plate (DQWP) compensator that dynamically reorients the input polarization to the instrument's eigenpolarization. Following installation of the DQWP, we measured an average increase in polarimetric efficiency of +17.5\% (to 87.4\%) and a reduction in instrumental polarization of -5.4\% (to 8.3\%) across all filters. Finally, we demonstrate the on-sky performance of the polarimeter with i' imaging of the debris disk around HR 4796, producing one of the closest inner-working-angle views of the bright, forward-scattering side of the disk. These results help pave the way for polarimeters on future extremely large telescopes such as GMT and ELT.
\end{abstract}


\section{Introduction}\label{sec:intro}

In the last two decades, large-diameter telescopes and technological advancements have enabled the direct imaging of circumstellar regions around nearby stars. These regions are comparable in scale to our own Solar-system ($\sim$\qty{100}{au} in extent) and have ushered in a new horizon of exoplanet research. By directly imaging exoplanets and their circumstellar environments, we can learn about their masses (via astrometry), their atmospheric composition (via spectroscopy), and their evolution (\citenum{sparks_imaging_2002,traub_direct_2010,bowler_imaging_2016,benisty_optical_2023}).

Detecting and characterizing circumstellar disks provide crucial insights into planet formation (\citenum{benisty_optical_2023}). As protoplanets form, they interact with their surrounding disks, shaping structures such as spirals, rings, gaps, and shadows (\citenum{andrews_observations_2020}). By directly imaging these features, we can study planet-disk interactions and infer the presence and properties of forming planets (\citenum{kley_planet-disk_2012,dong_how_2017,dong_what_2017}).

However, direct imaging of circumstellar disks is technically challenging. Resolving disk substructures requires high angular resolution. Substructure sizes are determined by the disk's pressure scale height, typically ranging from 10 to 100 au (\citenum{andrews_observations_2020,benisty_optical_2023}). Nearby star-forming regions are $\sim$150 pc away, which corresponds to angular resolutions finer than 70 mas. Achieving this resolution at visible and near-infrared wavelengths demands large-aperture telescopes equipped with extreme adaptive optics to correct for wavefront errors (\citenum{guyon_extreme_2018}).

Another challenge in direct imaging is the high contrast (relative brightness) between the star and the disk. In the visible and near-infrared, the star can be $\sim 10^{3}$ times brighter than the surrounding disk, making it difficult to extract the faint disk signal from the overwhelming stellar light (\citenum{benisty_optical_2023}). Diffraction control and post-processing algorithms are essential to suppress unwanted starlight and reveal the faint disk features.

Polarimetry provides a powerful tool to aid in direct imaging of disks by exploiting differences between the polarization of the disks and their host stars. While direct starlight is largely unpolarized, light scattered by sub-micron dust grains suspended in the disk is partially polarized ($\sim$10\%-80\%; \citenum{tazaki_effect_2019}). By capturing images in orthogonal polarization states and subtracting them, we can remove the unpolarized stellar light and isolate the polarized disk signal. This method, known as polarimetric differential imaging (PDI, \citenum{kuhn_imaging_2001}), suppresses stellar light by a factor of $10^{2}$ to $10^{4}$, outperforming conventional point spread function (PSF) subtraction techniques while also avoiding the inherent non-linear subtraction biases (\citenum{lafreniere_new_2007,soummer_detection_2012}). This makes modeling much simpler, as forward models can be directly compared to polarimetric images instead of requiring processing through the PSF subtraction algorithm (e.g., \citenum{monnier_multiple_2019,tschudi_quantitative_2021,ma_quantitative_2023}).

Polarimetric differential imaging requires precise calibration, as reflections within the telescope and instrument introduce unwanted polarization and mix incident polarization states (\citenum{tinbergen_accurate_2007}). While techniques exist to modulate the polarimetric signals and reduce these effects, they are imperfect--especially across broad wavelength ranges. Additionally, some instrumental polarization sources, such as the telescope's tertiary mirror (M3), cannot be mitigated through modulation alone because it is not feasible to add polarimetric optics upstream of the telescope mirrors. To correct for these effects, a model of the instrumental polarization must be fitted and applied to the measured data, ensuring an accurate extraction of the astrophysical polarimetric signal (\citenum{perrin_polarimetry_2015,holstein_polarimetric_2020,holstein_calibration_2020,zhang_characterizing_2023}).

In Spring of 2025, the Magellan Extreme Adaptive Optics instrument (MagAO-X; \citenum{males_magao-x_2020}), was upgraded to enable polarimetry (\citenum{souza_concept_2025}). In this paper, we will present an overview of the MagAO-X polarimeter (\autoref{sec:instrument-overview}), our efforts in calibrating the instruments (\autoref{sec:initial-characterization}), new hardware deployed for optimizing polarimetric efficiency (\autoref{sec:dqwp}), and the first on-sky disk images (\autoref{sec:on-sky-results}).

\section{Instrument Overview}\label{sec:instrument-overview}

MagAO-X is a visible-light (\qtyrange{500}{1050}{\nano\meter}) high-contrast instrument at the \qty{6.5}{\meter} Magellan Clay Telescope. It utilizes multiple wavefront sensors (WFS) and deformable mirrors (DM) at kilohertz speeds for extreme adaptive optics correction (\citenum{guyon_extreme_2018}). MagAO-X achieves a diffraction-limited resolution of \qty{19}{\mas} and has a field-of-view (FOV) of \ang{;;6}x\ang{;;6}, making it well-suited for circumstellar imaging of nearby stars. MagAO-X uses two electron-multiplication CCDs (EMCCDs), simultaneously, for dual-band (and polarimetric) imaging. The cameras are synchronized with an external trigger for contemporaneous acquisitions, utilizing different EM gains on each camera to optimize the signal-to-noise ratio (S/N).

\begin{figure}
    \centering
    \includegraphics[width=0.48\textwidth]{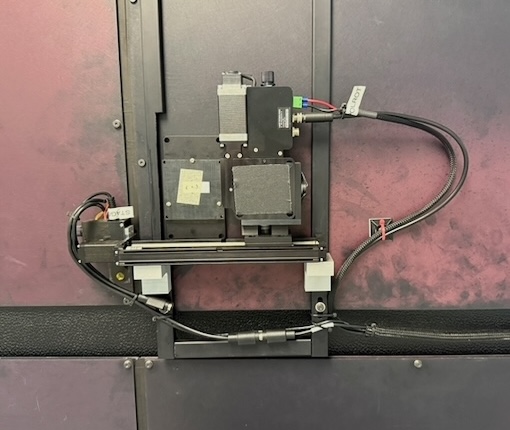}
    \includegraphics[width=0.48\textwidth]{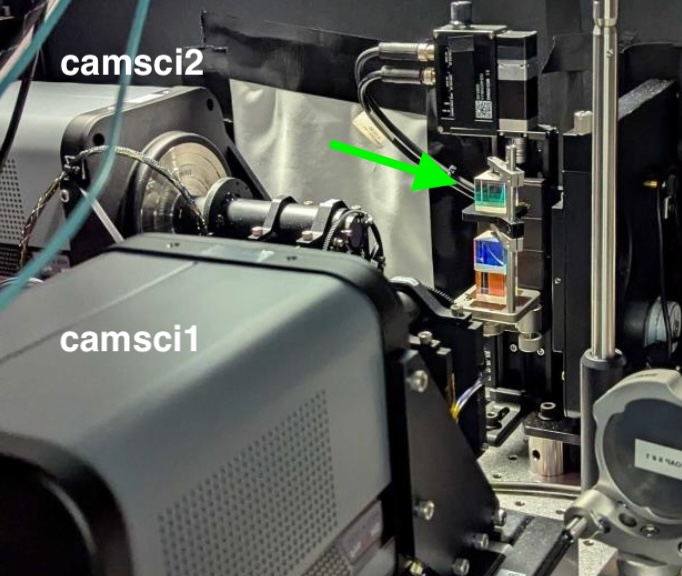}
    \caption{(Left) The HWP stage mounted on the exterior of the MagAO-X enclosure. (Right) The beamsplitter turret inside MagAO-X, hosting the PBS (top cube; green arrow).\label{fig:hwp}}
\end{figure}

The polarimetric optics in MagAO-X consist of: a broadband dielectric polarizing beamsplitter cube (PBS; Edmund-Optics \#49-871) and an achromatic half-wave plate (HWP; Edmund-Optics \#39-045), both shown in \autoref{fig:hwp}. According to manufacturing specifications, the PBS has an average extinction ratio of 500:1 (equivalent to a diattenuation of 99.8\%) across 700 nm to 1100 nm and an average throughput of 90\%. The 20 mm x 20 mm cube is installed in the f/69 converging beam upstream of the two science cameras and their respective filter wheels. The filters compatible with polarimetry are r', i', and z', as those are the only filters duplicated in both cameras' filter wheels. The HWP is a \qty{25.4}{\milli\meter} diameter achromat (\qty{650}{\nano\meter} to \qty{1100}{\nano\meter}).

The notable parts of MagAO-X that affect polarimetric performance are the telescope tertiary mirror (M3), the k-mirror image rotator (IMR), and the periscope between the two optical tables in the instrument (\citenum{tinbergen_accurate_2007,holstein_polarimetric_2020,zhang_characterizing_2023,mcintosh_characterizing_2026}). M3 causes instrumental polarization (IP; where unpolarized light becomes polarized) which is a nuisance to measure due to the mirror's size and location. The image rotator creates IP and elliptical retardance, which means incident polarized light is turned into a mixture of linear and circular polarization with different angles of linear polarization than before. M3 and the IMR both also move over the course of an observation, so these effects are dynamic. Finally, the optical periscope also acts as a rotated elliptical retarder.

\subsection{Differential Polarimetry}\label{sec:differential-polarimetry}

Differential polarimetry is the key for removing the unpolarized stellar signal (\citenum{kuhn_imaging_2001}). To explain, we denote the flux arriving to each camera as $f_1$ for \texttt{camsci1} and $f_2$ for \texttt{camsci2}. Because this light is split by a PBS, $f_1$ is horizontally polarized and $f_2$ is vertically polarized. This flux is the combination of (half) the incident unpolarized signal ($I_*$) and the inherent polarized signal. The polarized signal is made up of both the systematic instrumental polarization ($\epsilon_H$ and $\epsilon_V$) and the astrophysical polarized signal ($I_H$ and $I_V$),
\begin{align}
    f_1 = 0.5I_* + I_H + \epsilon_H \\
    f_2 = 0.5I_* + I_V + \epsilon_V.
\end{align}
Differencing these two fluxes removes the unpolarized signal completely
\begin{equation}
    \label{eqn:single-diff}
    f_2 - f_1 = (I_V - I_H) + (\epsilon_V - \epsilon_H).
\end{equation}
By definition, the difference of vertical and horizontal polarized intensities is the Stokes Q observable (\citenum{monnier_multiple_2019}). However, what remains in \autoref{eqn:single-diff} are the instrumental polarization terms ($\epsilon$). This signal is a fraction of incident stellar light (on the order of 5\% or more, depending on the instrument), which is orders of magnitude brighter than typical astrophysical signals. Another problem is that with a fixed PBS, we cannot measure Stokes U, which is the difference of +\ang{45} and \ang{-45} polarized light.

To solve both of these problems, we utilize the HWP to calculate the double-difference. We can use the HWP to arbitrarily reorient the incident angle of linear polarization. So, we can take horizontally polarized light and rotate it to the orthogonal, vertical orientation. This is how double-differencing removes the instrumental polarization--anything polarized downstream of the HWP will always produce the same systematic polarization, while the astrophysical signal is now in the orthogonal state
\begin{align}
    \left. f_2 - f_1\right|_{\theta=0^\circ} = (I_V - I_H) + (\epsilon_V - \epsilon_H) \\
    \left. f_2 - f_1\right|_{\theta=45^\circ} = (I_H - I_V) + (\epsilon_V - \epsilon_H).
\end{align}
By subtracting the intensities, we remove the systematic signal
\begin{equation}
\begin{aligned}
    \left. \left(f_2 - f_1\right)\right|_{\theta=0^\circ} - \left. \left(f_2 - f_1\right)\right|_{\theta=45^\circ} = \\
    \left[(I_V - I_H) - (I_H - I_V)\right] + \left[(\epsilon_V - \epsilon_H) - (\epsilon_V - \epsilon_H)\right] =\\
    2(I_V - I_H).
\end{aligned}
\end{equation}
To conserve flux, the double-difference is divided by two. To measure Stokes U, we use the HWP to reorient the incident $\pm$\ang{45} signal into horizontal or vertically polarized light (i.e., aligning it to the eigenpolarization of our instrument).

While we showed this example using an ideal beamsplitter, in reality there are transmission losses that are different for each beam. However, by double-differencing we also remove any bias in the polarized observable caused by the differential transmission.

\section{Polarimetric Characterization}\label{sec:initial-characterization}

\subsection{Polarization Generator}\label{sec:polarization-generator}

\begin{figure}
    \centering
    \includegraphics[width=0.48\textwidth]{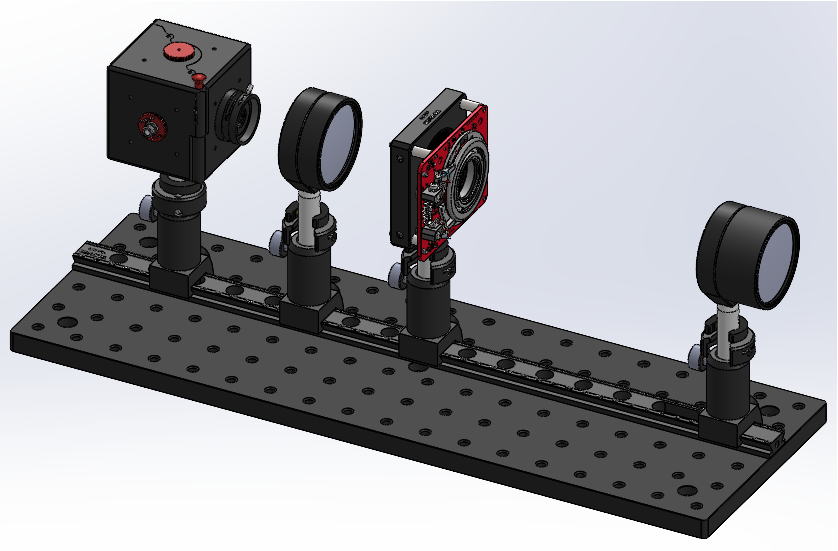}
    \includegraphics[width=0.48\textwidth]{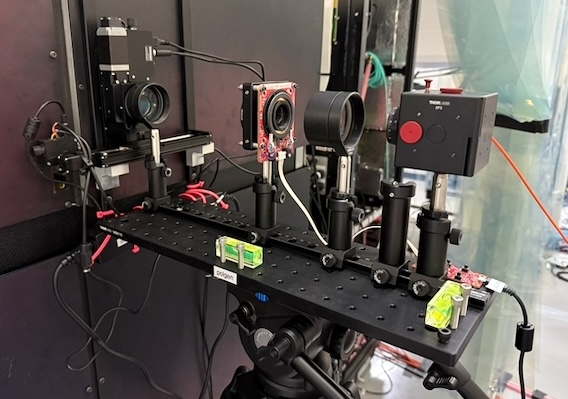}
    \caption{(Left) A computer model of the polarization generator. (Right) the polarization generator installed on its positioning tripod in front of MagAO-X in preparation for calibrating the instrument.\label{fig:polgen}}
\end{figure}

In order to measure the polarimetric effects of MagAO-X, we need to inject signal with a known polarization state. To accomplish this, we designed a polarization generator which generates effectively 100\% linearly polarized light with an arbitrary angle of linear polarization. The polarization generator consists of a Tungsten halogen lamp (Thorlabs \#SLS201L), a 50 mm integrating sphere (Thorlabs \#2P3), and a removable and remotely rotatable ultra-broadband dielectric linear polarizer (Meadowlark \#GPM-100-UNC). The polarizer is specified for the wavelength range \qty{300}{\nano\meter} to \qty{2700}{\nano\meter}, with an extinction ratio ${>}10^6$ (${>}$99.9999\% diattenuation) from \qty{450}{\nano\meter} to \qty{700}{\nano\meter}, and ${>}10^3$ (${>}$99.9\% diattenuation) for wavelengths longer than \qty{700}{\nano\meter}. The integrating sphere's output is reimaged with two achromatic doublet lenses to produce an f/11 focal ratio. This is so a single-mode fiber can be swapped in to create a polarized point source that matches the telescope's focal ratio. A model of the polarization generator is shown in \autoref{fig:polgen}.

The polarization generator must be set up outside of MagAO-X because MagAO-X's HWP is outside the enclosure (\citenum{souza_concept_2025}). To accomplish this, we mounted the optical components on a small optical table which mounts to a heavy-duty video tripod. The tripod is wheeled up to the MagAO-X instrument enclosure while it is off the telescope, and the tripod is leveled within \ang{0.1} of the MagAO-X optical bench.

\subsection{Data Acquisition}\label{sec:data-acquisition}

In November 2025 we used the polarization generator to measure the polarimetric response of MagAO-X. For each filter (r', i', and z'), we injected vertically polarized light and rotated both the HWP and the image rotator. The HWP was rotated between \ang{0} to \ang{90} in steps of \ang{11.25}, while the image rotator was rotated between \ang{-90} and \ang{0} in steps of \ang{11.25}. The image rotator angle is defined by the plane of reflection of its three constituent mirrors: at \ang{-90} the plane of reflection is horizontal and the second mirror is oriented to the right when looking into the beam. At \ang{0}, the second mirror is oriented vertically, and at \ang{90} it is to the left when looking into the beam.

Images are taken for each combination of HWP and image rotator angles (a total of 81 images). From these images, a dark frame is subtracted matching the exposure time and EM gain before sigma-clipping all pixel values with a cutoff of five standard deviations. The remaining pixels are summed to calculate the flux, and the flux error is estimated from the contribution of each pixel's individual photon noise and read noise to the sum. In general, the flux error is too small to affect our analysis--we are limited by systematic errors more than measurement uncertainty.

The normalized polarimetric observable is determined from the flux reflected by the PBS to \texttt{camsci2} (vertical) minus the flux transmitted to \texttt{camsci1} (horizontal) divided by the sum of the two fluxes--
\begin{equation}
    \label{eqn:stokes_observable}
    X = V - H;\quad I_X = V + H;\quad x = X / I_X.
\end{equation}
This observable ($x$) is equivalent to Stokes q in the camera's reference frame and has been normalized to a unitless quantity ranging from -1 (horizontal) to 1 (vertical). In order to measure Stokes U, the HWP is rotated by \ang{22.5} such that incident Stokes U light is transformed into the eigenpolarization of the instrument (i.e., vertical and horizontal), which is then measured by the two cameras. Note, we follow the IAU definition of Stokes parameters which means vertically polarized light is positive Stokes Q, and Stokes U is \ang{45} counter-clockwise from vertical when looking into the beam, and Stokes V is rotating counter-clockwise when looking into the beam \citenum{monnier_multiple_2019}.

\subsection{Mueller-Matrix Modeling}\label{sec:mueller-matrix-modeling}

Mueller-matrices offer a convenient way to describe the polarimetric response of intensity-based instruments. There is a proven history of fitting Mueller-matrices to astronomical polarimeters in order to correct for the non-ideal effects inherent to the complicated optical design of Nasmyth high-contrast instruments (\citenum{tinbergen_accurate_2007,perrin_polarimetry_2015,norris_vampires_2015,holstein_polarimetric_2020,joost_t_hart_full_2021,zhang_characterizing_2023,mcintosh_characterizing_2026}). When dealing with Stokes parameters, we can treat the quantites as a single vector, $\vec{S}=\left(I\quad Q\quad U\quad V \right)$. Using this formalism, we can describe any transformation of the Stokes vector as a matrix operation, defined as the Mueller matrix--
\begin{equation}
    \vec{S}^\prime = \mathbf{M}\cdot\vec{S}
    \equiv
    \begin{pmatrix}
    I\to I & Q\to I & U\to I & V\to I \\
    I\to Q & Q\to Q & U\to Q & V\to Q \\
    I\to U & Q\to U & U\to U & V\to U \\
    I\to V & Q\to V & U\to V & V\to V
    \end{pmatrix}\cdot
    \begin{pmatrix} I \\ Q \\ U \\ V \end{pmatrix}.
\end{equation}
In general, there are three aspects to a Mueller matrix: depolarization (creates unpolarized light from polarized light), diattenuation (creates polarized light from unpolarized light), and retardance (mixes polarization states). There is also the overall throughput, which is just a scalar multiplicative factor. Common optical components, like waveplates and polarizers, have formulae for generating their Mueller matrices, such that only one or two parameters are necessary to derive the full 16-element matrix (\autoref{sec:mueller-matrices}). Mueller matrices can be combined together using matrix multiplication from right to left
\begin{equation}
    \mathbf{M}_\text{comb} = \mathbf{M}_N\cdot ... \cdot\mathbf{M}_2\cdot\mathbf{M}_1.
\end{equation}
This is convenient as we can combine the models for multiple optical components into a single matrix as long as they are in the same reference frame.

Choosing the exact formulation of the MagAO-X Mueller-matrix model requires trade-offs; a more generic model may be more flexible (e.g., fitting every element of the Mueller matrix), but the increase in parameters can lead to difficulties optimizing the parameters and may lead to over-fitting. In general, we try to use the simplest model possible, only reaching for more complicated terms when needed to improve the fit. The MagAO-X model comprises the following components: (1) the telescope model, including M3, (2) HWP, (3) M4, which is a fold mirror between the HWP and IMR, (4) the IMR, (5) generic instrument term, and (6) PBS. 

The telescope term requires modeling M3 with appropriate rotation matrices to account for parallactic angle and telescope altitude. 
\begin{equation}
    \mathbf{M}_\text{Tel} = \mathbf{T}\left(a\right)\cdot \mathbf{M}_\text{M3}\cdot\mathbf{T}\left(p\right),
\end{equation}
where $\mathbf{T}$ is a rotation matrix, $a$ is the telescope altitude\footnote{MagAO-X is nominally on the East Nasmyth platform (NASE), but in S26A MagAO-X was installed on the West Nasmyth platform (NASW), so for the on-sky data from that semester we actually used $\mathbf{T}\left(-a\right)$ to account for the different geometry.}, and $p$ is the parallactic angle. The HWP is modeled as a linear retarder without diattenuation. The M4 is modeled as a linear retarder with linear diattenuation. The IMR is modeled as an elliptical retarder with linear diattenuation. The generic instrument term, which is mostly affected by the optical periscope, is modeled as an elliptical retarder with linear diattenuation. Finally, the PBS is modeled as a Wollaston prism with different linear diattenuations for each camera. 

The overall instrument model as a function of instrument state ($\mathbf{\Theta}$) and hyperparameters ($\mathbf{\Omega}$) is
\begin{equation}
    \mathbf{M}\left(\mathbf\Theta \left| \mathbf\Omega\right.\right) =
    \mathbf{M}_\text{PBS}
    \cdot\mathbf{M}_\text{Inst}
    \cdot\mathbf{M}_\text{IMR}
    \cdot\mathbf{M}_\text{M4}
    \cdot\mathbf{M}_\text{HWP}
    \cdot\mathbf{M}_\text{Tel}.
\end{equation}
The full mathematical formulation for each individual component is described in \autoref{sec:magaox-mueller-matrices}. We do not fit the telescope term ($\mathbf{M}_\text{Tel}$) because it is unable to be measured with the polarization generator. We do not fit the polarization generator's diattenuation as it has an extinction ratio greater than $10^3$ at all wavelengths (and greater than $10^6$ for the r' filter), which is greater than the extinction of the PBS. All the parameters and hyperparameters for each component are itemized in \autoref{tbl:parameters}.

\begin{table}
    \centering
    \caption{MagAO-X Mueller-matrix model parameters. $\theta$ represents an angle, while $\delta$ represents an offset to a given angle (e.g., $\theta_\text{HWP}$ is the value from the instrument telemetry with a fitted offset, $\delta_\text{HWP}$). $\chi$ is diattenuation and $\eta$ is retardance--for elliptical retarders we specify the retardance for each orthogonal polarization with a superscript.\label{tbl:parameters}}
    \begin{tabular}{clll}
    \hline\hline
    Term & State parameters ($\mathbf\Theta$) & Fit parameters ($\mathbf\Omega$) & Notes \\
    \hline
    $\mathbf{M}_\text{Tel}$ & $p$, $a$ & $\chi_\text{M3}$, $\eta_\text{M3}$ & Not used for internal fits; only used on-sky \\
    $\mathbf{M}_\text{HWP}$ & $\theta_\text{HWP}$ & $\delta_\text{HWP}$, $\eta_\text{HWP}$ & \\
    $\mathbf{M}_\text{M4}$ & & $\chi_\text{M4}$, $\eta_\text{M4}$ & Simple mirror model \\
    $\mathbf{M}_\text{IMR}$ & $\theta_\text{IMR}$ & $\delta_\text{IMR}$, $\chi_\text{IMR}$, $\eta^Q_\text{IMR}$, $\eta^U_\text{IMR}$, $\eta^V_\text{IMR}$& \\
    $\mathbf{M}_\text{Inst}$ & & $\theta_\text{Inst}$, $\chi_\text{Inst}$, $\eta^Q_\text{Inst}$, $\eta^U_\text{Inst}$, $\eta^V_\text{Inst}$ & Dominated by optical periscope\\
    $\mathbf{M}_\text{PBS}$ & camera & $\chi_1$, $\chi_2$ & Separate diattenuation for each camera \\
    \end{tabular}
\end{table}

To apply this model, we use the FITS headers (based off MagAO-X's telemetry) for the instrument state (i.e., the HWP and IMR angles) along with a vector of parameters mapping to the tunable hyperparameters. This generates a single Mueller matrix, from which we extract the observation vector $\vec{w}$ (the first row of the matrix mapping the four Stokes parameters to intensity, $X\rightarrow I$, which is recorded by the detector).
\begin{equation}
\vec{w} = \begin{pmatrix}
1 & 0 & 0 & 0 \\
\end{pmatrix}\cdot\mathbf{M}.
\end{equation}
Since we calculate the single-difference and double-difference of our data, we perform the same operations on the observation vectors. Then, from each single- or double-difference observation vector, we assemble the full data observation matrix by vertically stacking the vectors
\begin{equation}
    \mathbf{W} = \begin{pmatrix}
        \vec{w}_1 \\ \vec{w}_2 \\ \vdots \\ \vec{w}_n
    \end{pmatrix}.
\end{equation}

\begin{figure}
    \centering
    \includegraphics[width=0.8\textwidth]{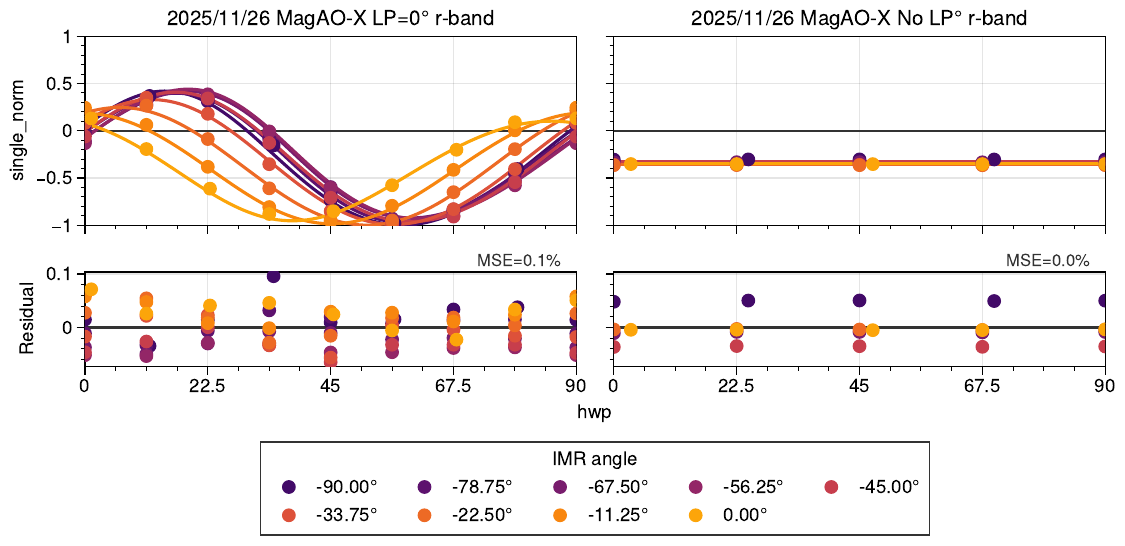}
    \includegraphics[width=0.8\textwidth]{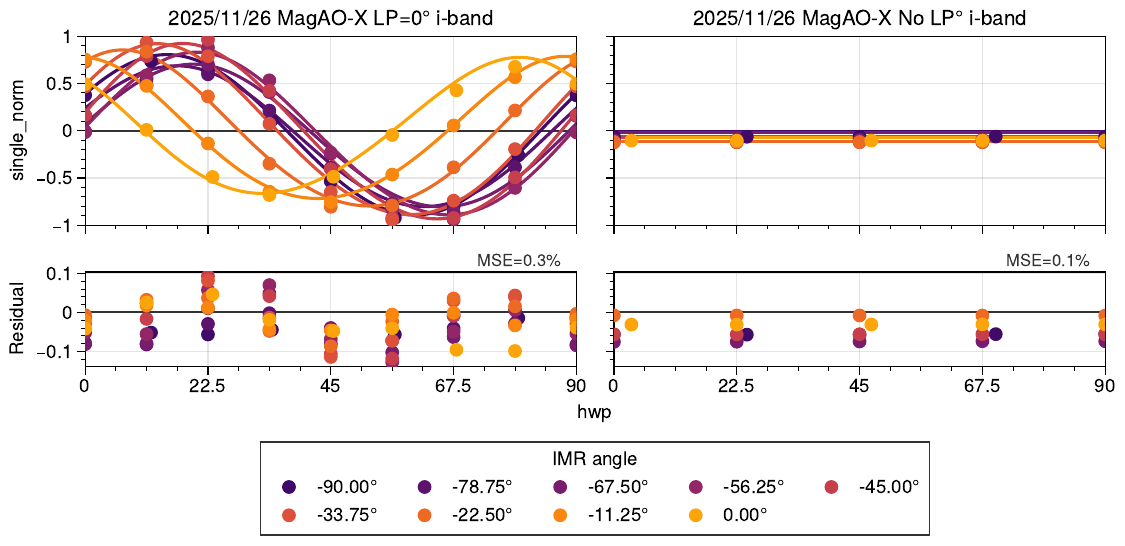}
    \includegraphics[width=0.8\textwidth]{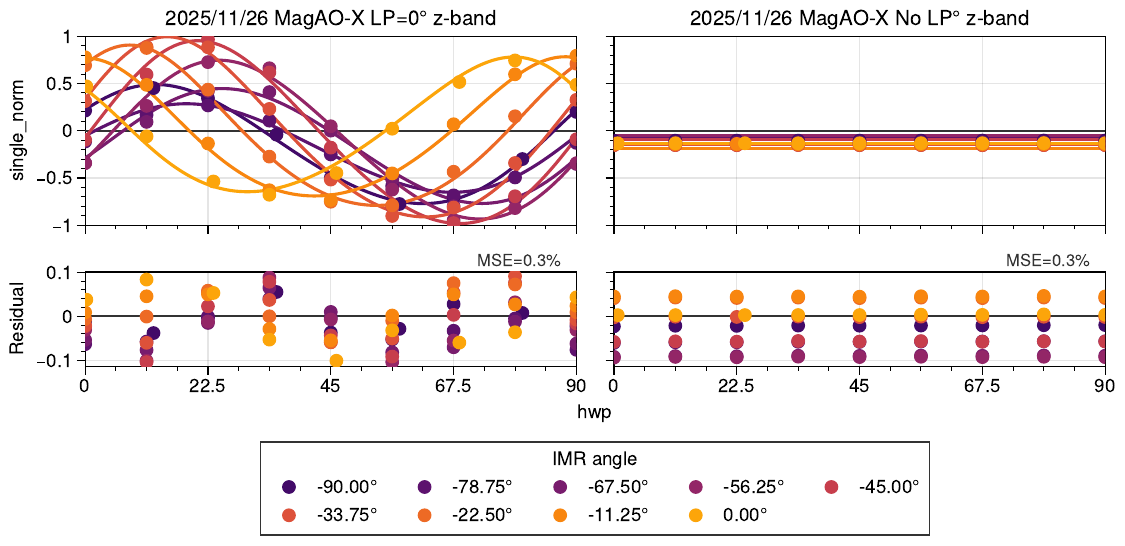}
    \caption{The normalized single-difference fluxes alongside the optimized Mueller-matrix model. The left column shows data with polarized light injected by the polarization generator, while the right column is unpolarized light. Each row corresponds to the three observing filters (r', i', and z') as well as showing the residuals between the model and the data.\label{fig:single-diffs-pre-qwp-model}}
\end{figure}

For modeling, we extract the flux from $\mathbf{W}$ by indexing the first column
\begin{equation}
    \hat{y} = \mathbf{W}\cdot\begin{pmatrix}
        1 & 0 & 0 & 0
    \end{pmatrix}^T.
\end{equation}
We optimize for our hyperparameters by minimizing mean-squared error (MSE) against the measured single-difference fluxes,
\begin{equation}
    MSE = \frac{1}{N}\sum_i^N{\left(y_i - \hat{y}_i \right)^2}.
\end{equation}
For this work, we used a Nelder-Mead numerical optimizer to estimate the best-fitting parameters using conservative bounds on the parameters. We performed a joint fit to the normalized single-difference fluxes with and without the polarizer. Using both data products increases the leverage the model has for fitting diattenuation, since any non-zero flux without the polarizer comes from diattenuation. The models, with residuals, are shown in \autoref{fig:single-diffs-pre-qwp-model}.

Because we injected vertically polarized light, an ideal polarimeter would have a pure sinusoidal response from +1 to -1 with a period of \ang{90} as the HWP rotates without any modulation due to the image rotator. As shown, though, there are inefficiencies which compress the curves and retardance due to the image rotator. The r' data are especially concerning--there is a strong asymmetry in the polarimetric response, going from -1 to +0.3. We attribute this to the PBS--the manufacturer's transmission and reflection curves show significant (${>}$10\%) leakage of vertically polarized light below \qty{650}{\nano\meter}, which is roughly half the r' bandpass. As flux is conserved, the leakage of the vertically polarized light means less vertical light is reflected to \texttt{camsci2}, causing the asymmetric response.


\subsection{Results}\label{sec:initial-results}

\begin{figure}
    \centering
    \includegraphics[width=0.48\textwidth]{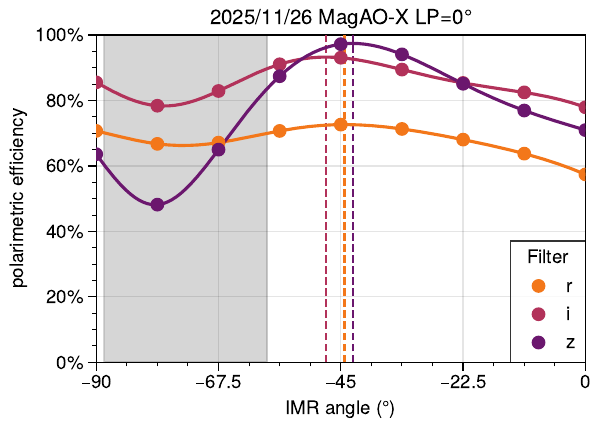}
    \includegraphics[width=0.48\textwidth]{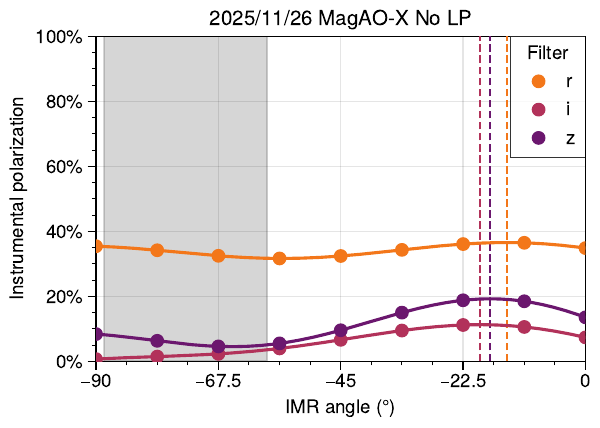}
    \caption{(Left) polarimetric efficiency for each filter as a function of IMR angle based on the Mueller-matrix model excluding M3. (Right) instrumental polarization for each filter as a function of IMR angle based on the Mueller-matrix model excluding M3. In both, the gray region indicates the range of IMR angles used on sky.\label{fig:pol-eff-pre-qwp}}
\end{figure}

From our Mueller-matrix model we can learn a few things about MagAO-X as a polarimeter, namely the polarimetric efficiency and the total instrumental polarization. The polarimetric efficiency is the total amount of linear polarization measured as a fraction of incident linear polarization--note that this makes no distinction about the angle of linear polarization, since we can arbitrarily rotate the output Stokes vector without loss of flux. Therefore, the efficiency is calculated from the Mueller-matrices for the double-difference of Stokes Q and U (i.e., at the HWP angles \ang{0}, \ang{45}, \ang{22.5}, and \ang{67.5}), for each IMR angle. From the two matrices, the efficiency is the average combination of the Q$\rightarrow$Q, U$\rightarrow$U, Q$\rightarrow$U, and U$\rightarrow$Q terms (\citenum{holstein_polarimetric_2020}). This way, we account for all linear polarized signal regardless of the angle of linear polarization. We can also measure the instrumental polarization as the fraction of linear polarized light generated when unpolarized light is injected into the system. We plot the polarimetric efficiency and instrumental polarization for each filter in \autoref{fig:pol-eff-pre-qwp}. These plots are shaded to show the range of IMR angles used on sky according to the MagAO-X pupil offset.

Immediately we can see that the polarimeter is sub-optimal: the average polarimetric efficiency is strongly constrained by the image rotator angle, and due to the MagAO-X pupil offset we are operating in a regime with poor efficiency. The average efficiency within the shaded region for the r' filter is 68\%, 82\% for i', and 60\% for z'. We also see that the r' efficiency has a much lower peak efficiency, which we attribute to the leakage of the PBS. Looking at the instrumental polarization, the average within the shaded region is 33\% for r', 2\% for i', and 6\% for z'.

\begin{figure}
    \centering
    \includegraphics[width=0.6\textwidth]{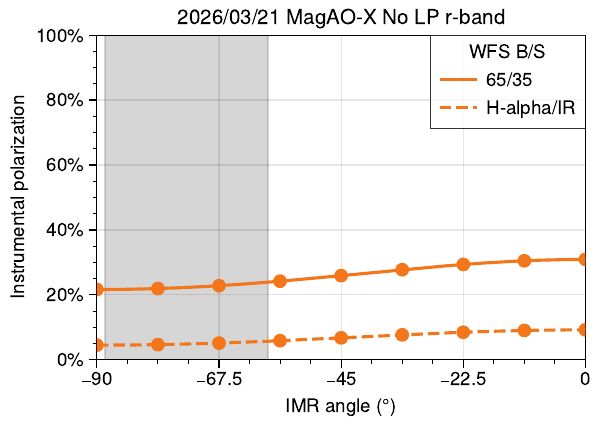}
    \caption{Comparing the r' instrumental polarization based on models fit to calibration data injected with unpolarized light. The solid curve uses the 65/35 gray WFS beamsplitter, and the dashed curve uses the H-alpha/IR WFS beamsplitter. These data were taken in 2026, following the installation of a polarimetric compensator (\autoref{sec:dqwp}). The gray region marks the range of IMR angles used on sky. The average instrumental polarization within the gray region is 22\% and 5\%, respectively. \label{fig:ip-wfs-bs}}
\end{figure}

The large instrumental polarization at r' appears to be from the WFS beamsplitter. There are two such beamsplitters with a \ang{45} angle-of-incidence (AOI): a 65\%/35\% gray beamsplitter with a cut-on wavelength around \qty{600}{\nano\meter} (the middle of the r' filter), and a dichroic beamsplitter with a critical wavelength of \qty{740}{\nano\meter}. The source of the r' instrumental polarization is confirmed in \autoref{fig:ip-wfs-bs}, which shows a large reduction in IP when switching to the dichroic beamsplitter.

To improve the performance at r', we plan to install a new PBS in November 2026, after which we will recharacterize the performance of the polarimeter. To improve the instrumental polarization we could install a new WFS beamsplitter, however this is a significant instrument modification and is not planned.

One way to improve the polarimetric efficiency for all filters would be to change the MagAO-X pupil offset such that we operate in a more favorable range of IMR angles (around \ang{0}; \autoref{fig:pol-eff-pre-qwp}), but this has two major drawbacks. First, changing the pupil offset would require a major realignment of the instrument's telescope simulator, pupil apodizing masks, and Lyot stops. Second, this does not address the dynamic nature of the IMR effects. In the next section, we will describe our solution for active compensation of the crosstalk within the instrument.

\section{Dual Rotating Quarter-Wave Plate Compensator}\label{sec:dqwp}

To address the polarimetric inefficiencies dynamically we present a dual rotating quarter-wave plate compensator (DQWP). This technique has been demonstrated before on SCExAO/VAMPIRES (\citenum{norris_vampires_2015,lucas_visible-light_2024}) to arbitrarily correct for crosstalk and to keep the angle of linear polarization stabilized to the instrument reference frame. In simple terms, this means if we inject vertically polarized light into MagAO-X, we measure vertically polarized light with maximum efficiency at the detectors.

The DQWP is installed in a collimated beam on the upper bench of MagAO-X, after the tweeter DM and prior to the upper periscope mirror. We use two zero-order QWPs with a design wavelength of \qty{780}{\nano\meter} (Meadowlark \#NQM-200-0780), as our simulations showed this could work better than a super-achromat while also being less expensive. The QWPs are mounted within two rotation mounts which enable continuous \ang{360} range of motion (Thorlabs \#ELL21). The entire module is on a linear stage so it can be removed during non-polarimetric observations (Zaber \#X-LSM100A-E03).

\subsection{Calibrating the DQWP}\label{sec:dqwp-calibration}

\begin{figure}
    \centering
    \includegraphics[width=\textwidth]{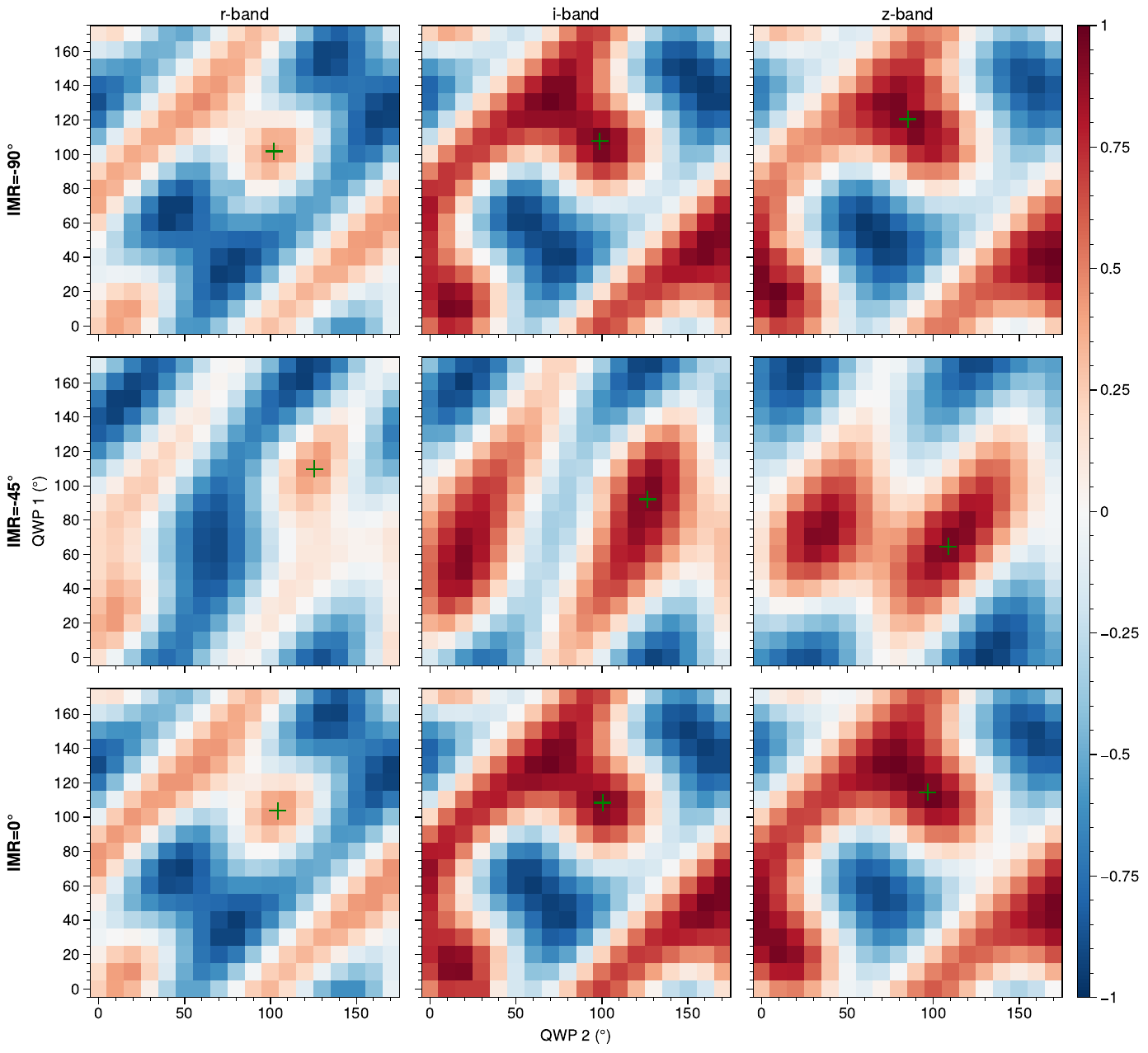}
    \caption{Surfaces of normalized polarimetric flux (from -1 to 1) as a function of QWP angles. The surfaces are shown with a diverging colormap and no interpolation. Each column corresponds to each observing filter, and each row corresponds to a different IMR angle. The maxima used for the tracking law are shown with green crosses.\label{fig:qwp-surfaces}}
\end{figure}

The DQWP is calibrated by injecting vertically polarized light with the polarization generator and performing a grid search on each QWP from \ang{0} to \ang{180} in steps of \ang{10}. For each QWP angle pair, we calculated the normalized Stokes Q via single-difference--when this quantity is maximized we have both maximized the polarimetric efficiency and reoriented the eigenpolarization of the instrument to match the PBS. We perform this grid search as a function of IMR angle from \ang{-90} to \ang{0} in steps of \ang{5} and for each observing filter. A surface map of the measured normalized polarimetric flux for a few IMR angles is shown in \autoref{fig:qwp-surfaces}.

\begin{figure}
    \centering
    \includegraphics[width=\textwidth]{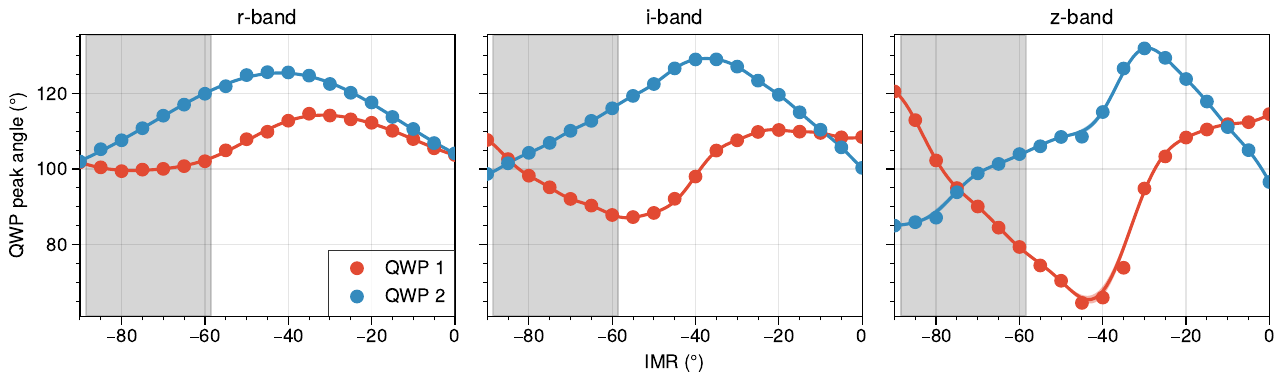}
    \caption{The optimal QWP angles as a function of IMR angle extracted from the normalized flux surfaces (\autoref{fig:qwp-surfaces}). The data points are shown in red for the first QWP and blue for the second QWP, along with solid lines representing the RBF fitted to each. Each column corresponds to the different observing filters. The gray region denotes the range of IMR angles used on sky.\label{fig:qwp-opt-angles}}
\end{figure}

The resulting surfaces show multiple peaks, although we focus only on the peaks which are consistently maxima and require the least motion of the QWPs throughout the IMR angle range. The peaks are estimated by fitting a bivariate quadratic function to the flux surfaces starting from a manual initial guess. We fit two radial-basis-function (RBF) Gaussian processes to interpolate the optimal QWP angle pairs. The extracted angles and RBF functions are shown in \autoref{fig:qwp-opt-angles}. These functions are deployed as a tracking law such that the two QWPs respond to any change in the IMR angle or observing filters.

\subsection{Polarimetric Characterization and Results}\label{sec:post-dqwp-characterization}

\begin{figure}
    \centering
    \includegraphics[width=0.8\textwidth]{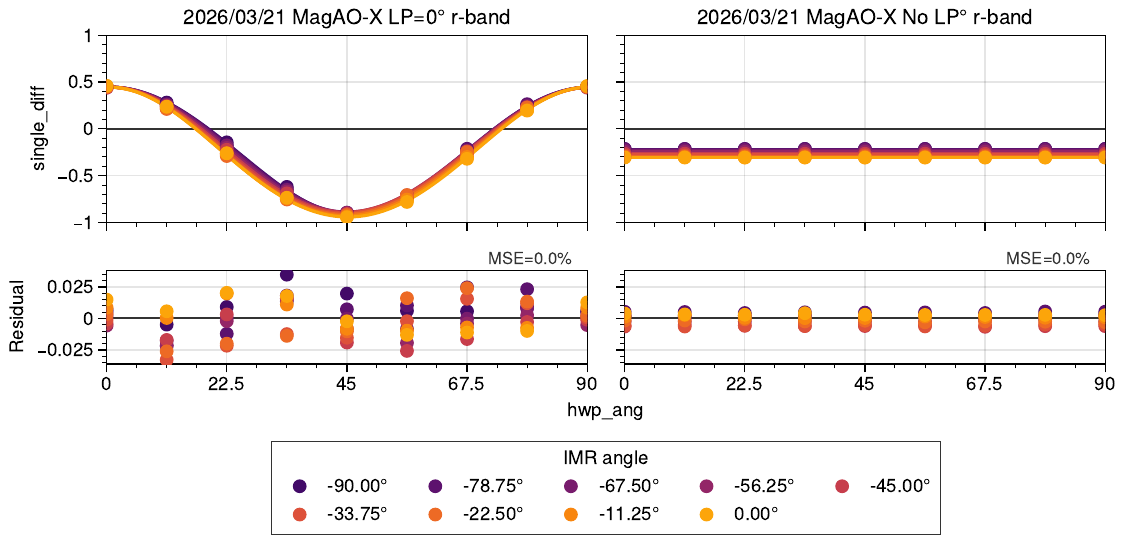}
    \includegraphics[width=0.8\textwidth]{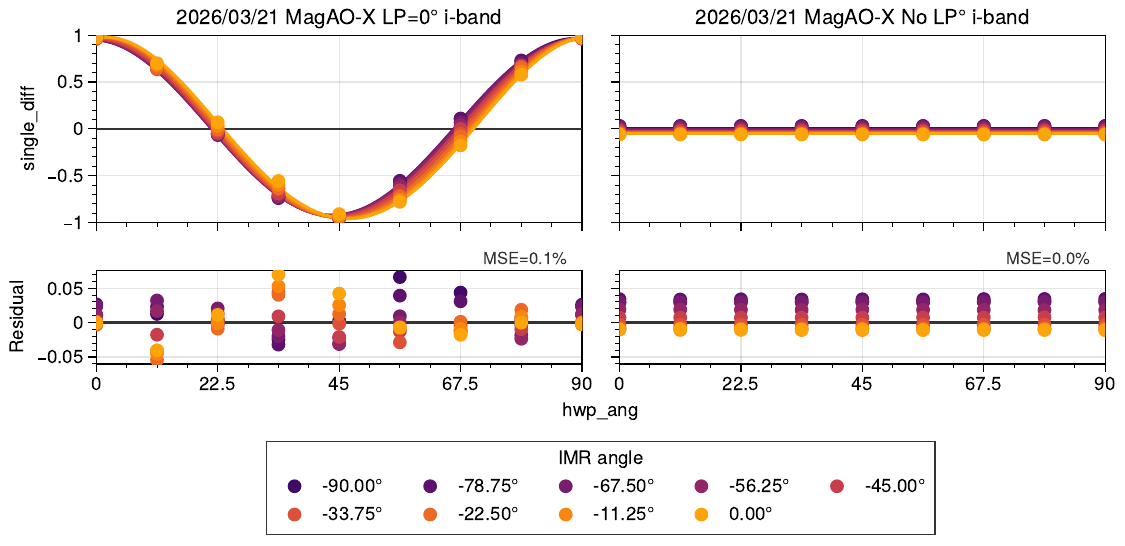}
    \includegraphics[width=0.8\textwidth]{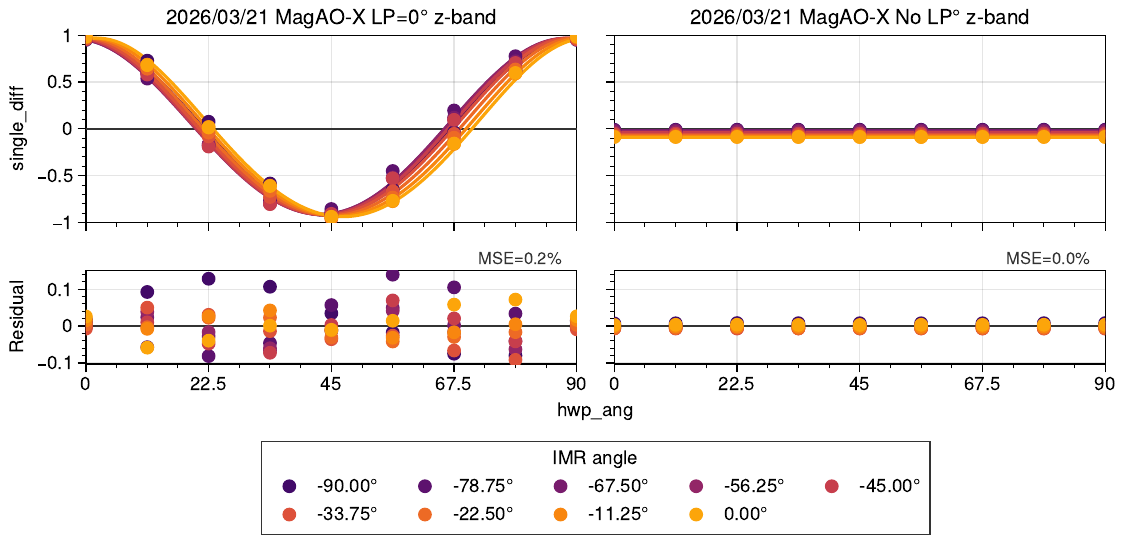}
    \caption{The same as \autoref{fig:single-diffs-pre-qwp-model} following the deployment of the DQWP compensator.\label{fig:single-diffs-post-qwp-model}}
\end{figure}

We performed another characterization of the polarimeter with the polarization generator using the DQWP, allowing the two QWPs to settle to their optimal angles when we changed the IMR angle. As before, we calculated the normalized single differences and fitted our model using a numerical optimizer. We used the same model, but since the DQWP should be correcting for the effects of the IMR, we simply changed our initial guess for the retardance of the IMR to be 0 waves for the three retardances in the IMR matrix. Adding additional terms to account for each QWP as a linear or elliptical retarder resulted in worse residuals overall, although more effort studying the model tradeoffs is necessary (\autoref{fig:single-diffs-post-qwp-model}).

\begin{figure}
    \centering
    \includegraphics[width=0.48\textwidth]{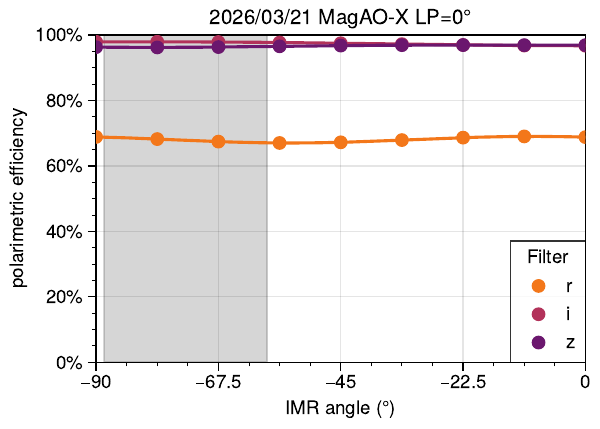}
    \includegraphics[width=0.48\textwidth]{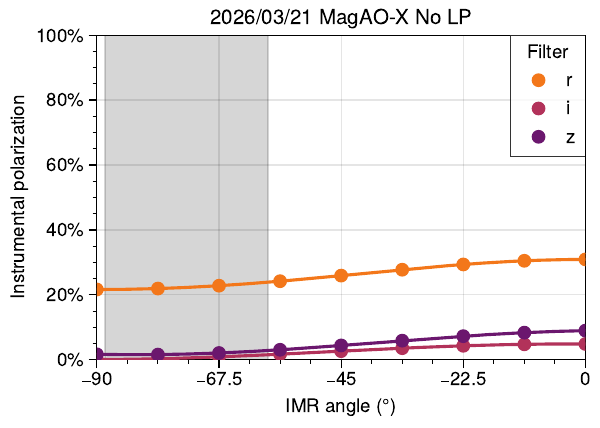}
    \caption{The same as \autoref{fig:pol-eff-pre-qwp} following the deployment of the DQWP compensator.\label{fig:pol-eff-post-qwp}}
\end{figure}

We show the polarimetric efficiency and instrumental polarization with the DQWP in \autoref{fig:pol-eff-post-qwp}. The compensator clearly increases the polarimetric efficiency and removes almost all modulation induced by the IMR. The imperfections are likely due to sub-optimal orientation of the two QWPs using our control law. The r' filter is still significantly worse performing than i' or z' due to the PBS. Interestingly, the instrumental polarization has also decreased. This means that previously the instrumental polarization from different sources was constructively adding together, but now some of it is fortuitously cancelling out. We tabulated the average polarimetric efficiency and instrumental polarization before and after the installation of the DQWP in \autoref{tbl:pol-eff}.

\begin{table}
    \centering
    \caption{MagAO-X polarimetric efficiency and instrumental polarization. All values are averaged over the IMR angle range used on sky.\label{tbl:pol-eff}}
    \begin{tabular}{ccccccc}
    \hline\hline
           & \multicolumn{3}{c}{Polarimetric Efficiency} & \multicolumn{3}{c}{Instrumental Polarization} \\
    Filter & Pre-DQWP & Post-DQWP & (+/-) & Pre-DQWP & Post-DQWP & (+/-) \\
    \hline
    r' & 67.6\% & 67.9\% & +0.3\% & 33.3\% & 22.4\% & -10.9\%\\
    i' & 82.0\% & 97.9\% & +15.9\% & 1.9\% & 0.6\% &  -1.3\% \\
    z' & 59.9\% & 96.3\% & +36.4\% & 5.8\% & 1.9\% & -3.9\% \\
    ave. & 70.0\% & 87.4\% & +17.5\% & 13.7\% & 8.3\% & -5.4\% \\
    \end{tabular}
\end{table}

\subsection{Optical Effects Caused by the DQWP}

As mentioned, the DQWP is in a collimated beam following the tweeter DM, and is not in pupil plane. This means each QWP shifts the pupil image and the PSF due to imperfect parallelism. We specifically purchased high-quality optics which have $<$\qty{1}{arcmin} parallelism, however this is still significant. It does not matter much that the PSF moves, since this is sensed and corrected by the AO loop; what is concerning is the shift in the pupil image.

The interaction matrix between the WFS and DM maps from WFS detector pixels to DM actuators, so any shift of the pupil image on the WFS will cut off high-spatial-frequency correction from the AO loop. Normally, MagAO-X uses a pupil-tracking loop to keep the pupil images aligned within 0.1 pixels (0.05 actuators). However, specifically because the DQWP is between the WFS and DM, we cannot use this loop to correct for misalignments because we are not shifting just the pupil image, but rather the entire interaction matrix. Furthermore, we cannot use a static offset because the two QWP angles change over the course of an observation. In practice, we measured shifts on the order of 0.5 pixels (0.25 actuators) when rotating the QWPs \ang{180}, which is significant for high-contrast imaging at visible wavelengths. Future work is planned to measure the shifts caused by each QWP, individually, to create a lookup table for offsetting the pupil-alignment loop when using the DQWP.

\section{On-Sky Polarimetric Imaging of A Circumstellar Disk}\label{sec:on-sky-results}

To prove the efficacy of the polarimeter, we observed known debris disk host HR 4796 (\citenum{hinkley_speckle_2009,perrin_polarimetry_2015,milli_near-infrared_2017,milli_optical_2019}).  We observed HR 4796 on 2026/05/13 (UT) for $\sim$\qty{2}{\hour} in good conditions. We used the i' filter since it has the highest polarimetric efficiency. We used the small Lyot coronagraph (inner working angle: \qty{63}{mas}) and utilized the DM to place fiducial speckles for astrometric alignment (\citenum{sahoo_precision_2020}). We used a detector integration time of \qty{0.23}{\second} with EM gain of \num{150} on \texttt{camsci1} and \num{180} for \texttt{camsci2}. We used the DQWP compensator and switched the HWP between \ang{0}, \ang{45}, \ang{22.5}, and \ang{67.5} every \qty{30}{\second}. The total integration time was \qty{73}{\minute} and the total field rotation was \ang{92}.

During the observation we noticed significant residual atmospheric dispersion, which smears the PSF in the focal plane leading to leakage around the coronagraph mask. This smearing is not static, changing with the telescope elevation. We also noticed that when we rotated the QWPs, the dispersion would change slightly. However, the majority of the residual dispersion appears to be due to the atmospheric dispersion corrector's (ADC) control law. Nonetheless, active ADC control (\citenum{twitchell_dynamic_2026}) would improve the performance of the polarimeter without requiring significant calibration or lookup tables for using the ADC with the DQWP.

\subsection{Data Processing}\label{sec:data-processing}

The data were reduced by subtracting a dark frame matching the detector integration time and EM gain. We converted the data to units of e-/s using the camera conversion gain, EM gain, and integration time. Then, we measured the location of each of the four calibration speckles and estimated the location of the star from the intersection of the cross formed by the points (\citenum{savransky_computer_2013}). We shifted the data so the star is centered in the image and cropped the image to 512x512 pixels. Because MagAO-X has a dual-beam camera setup, there are non-common path differences in the plate scale and position angle offset which can affect the differential imaging procedure, so we utilized the calibration speckle locations to fit an affine transform between the two cameras and applied it to the \texttt{camsci2} data.

Using the registered data we created difference images by subtracting the \texttt{camsci1} frame from the \texttt{camsci2} frame (\texttt{camsci2}$-$\texttt{camsci1}). We also created the sum image by adding the two frames together. We then grouped single-difference frames into cycles with HWP angles \ang{0}, \ang{45}, \ang{22.5}, and \ang{67.5} for double-differencing, making sure that each cycle occurred within a span of \qty{120}{\second} with a maximum parallactic angle rotation between frames of \ang{1}. After double-differencing we had \num{4535} individual Stokes cubes comprising of I$_\mathrm{Q}$, I$_\mathrm{U}$, Q, and U images.

Then, we calculated the double-difference Mueller-matrix using our fitted model. Since we cannot fit the M3 model with the polarization generator, for $\mathbf{M}_\text{M3}$ we used an idealized mirror model based on the refractive index of silver. Future work is planned to actually fit the M3 Mueller-matrix using polarized and unpolarized standard stars.

We performed least-squares Mueller-matrix inversion (\citenum{perrin_polarimetry_2015,holstein_polarimetric_2020}) for each of our double-difference images to estimate the calibrated Stokes images. From the calibrated Stokes images, we performed ad-hoc instrumental polarization removal by measuring the average stellar polarized flux within the (partially transmissive) coronagraph mask. The residual flux was removed by subtracting a scaled version of the Stokes I image (e.g., \citenum{follette_seeds_2015,avenhaus_disks_2018,de_regt_polarimetric_2024})--
\begin{equation}
    X' = X - c_X I_X,
\end{equation}
where $X$ is Stokes Q or U, $c_X$ is the average stellar polarized flux, and $I_X$ is the Stokes I image. Then, we coadded the calibrated images using an outlier-resilient mean. Because we had \num{4535} images, we performed the coadding in chunks, using memory-mapping to only read \qty{64}{rows} of pixels at a time. This kept the routine tractable on a laptop with \qty{64}{\giga\byte} of RAM.

\subsection{Stokes Images}\label{sec:stokes-images}

\begin{figure}
    \centering
    \includegraphics[width=0.48\textwidth]{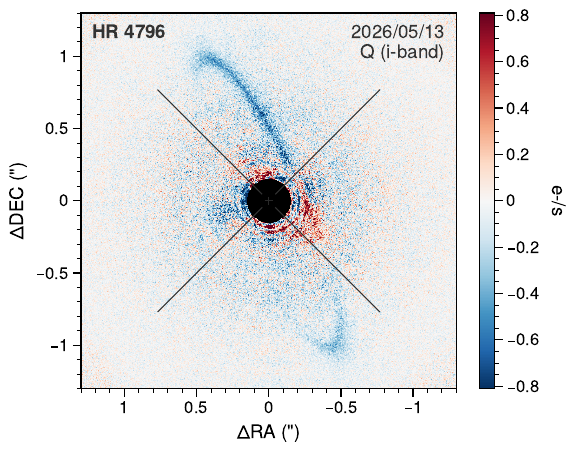}
    \includegraphics[width=0.48\textwidth]{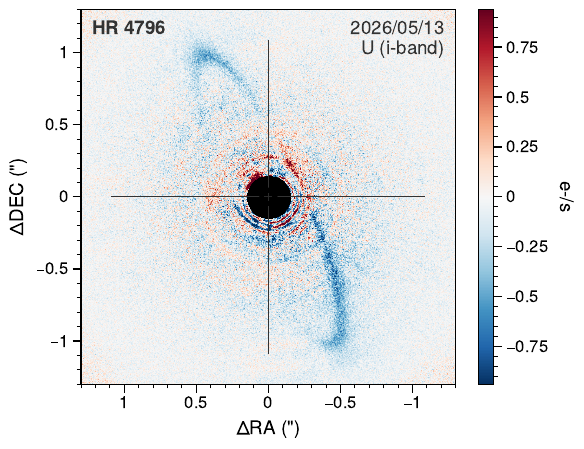}
    \caption{Stokes Q (left) and Stokes U (right) images of HR 4796 shown with linear, diverging colormaps. Both images have crosses showing the null angles of linear polarization. A black circle covers both the coronagraph mask and the bright residual stellar speckles.\label{fig:hr4796_q_u}}
\end{figure}

\begin{figure}
    \centering
    \includegraphics[width=0.48\textwidth]{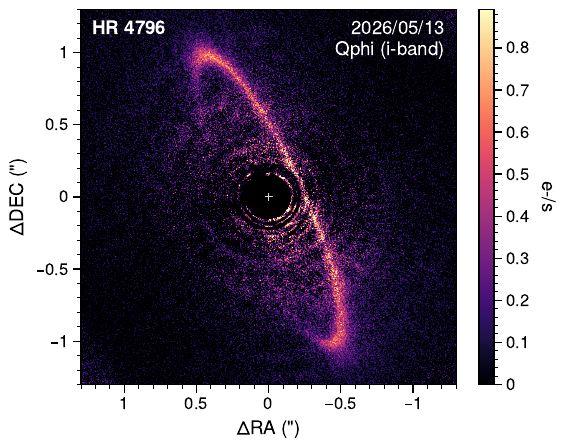}
    \includegraphics[width=0.48\textwidth]{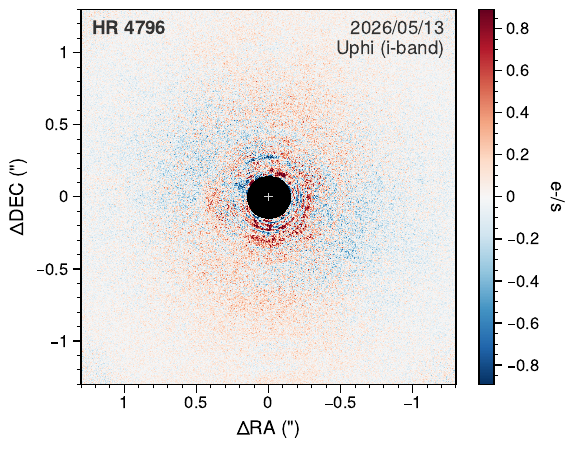}
    \caption{Stokes Q$\mathrm\phi$ (left) and Stokes U$\mathrm\phi$ (right) images of HR 4796 shown with linear colormaps. The Q$\mathrm\phi$ image is effectively the polarized intensity, while the U$\mathrm\phi$ image (shown with a diverging colormap) is the effective polarized intensity error. The U$\mathrm\phi$ image is scaled to the same maxima as the Q$\mathrm\phi$ image. A black circle covers the coronagraph mask and the bright residual stellar speckles.\label{fig:hr4796_qphi_uphi}}
\end{figure}

We show the final Stokes Q and U images in \autoref{fig:hr4796_q_u}. Due to the centrosymmetric nature of the scattering geometry, we calculated the azimuthal Stokes Q$\mathrm\phi$ and U$\mathrm\phi$ to produce a polarized intensity and polarized noise image (\citenum{schmid_limb_2006,follette_seeds_2015,avenhaus_disks_2018,monnier_multiple_2019})--
\begin{align}
    Q\phi = -Q\cos{2\theta} - U\sin{2\theta} \\
    U\phi =  Q\sin{2\theta} - U\cos{2\theta},
\end{align}
where $\theta$ is the azimuthal angle East of North for each pixel. We utilize a common technique of optimizing the Q$\mathrm\phi$ image by offsetting $\theta$ to minimize the absolute U$\mathrm\phi$ flux around the disk (e.g., \citenum{follette_seeds_2015,avenhaus_disks_2018,de_regt_polarimetric_2024}). The optimized offset angle was \ang{1.9}. We show Q$\mathrm\phi$ and U$\mathrm\phi$ in \autoref{fig:hr4796_qphi_uphi}.

The Stokes Q and U images show the characteristic ``butterfly'' pattern expected from the centrosymmetric single-scattering geometry of a circumstellar disk. In all the images, there are still bright residual speckles from the star, both around the coronagraph mask and further out, where the passive DM speckles created by the diffractive gridding of the tweeter DM lie. We believe this is at least partially due to the suboptimal atmospheric dispersion control. Nonetheless, this result is one of the closest inner working angle images of this disk, resolving the bright, forward-scattering, near side of the disk with minimal contamination from the star.

\section{Conclusions}\label{sec:conclusions}

In this work we have done the following:
\begin{itemize}
    \item Described the optical setup of the MagAO-X polarimeter (\autoref{sec:instrument-overview})
    \item Built a polarization generator for injecting broadband polarized and unpolarized light (\autoref{sec:polarization-generator})
    \item Used the polarization generator to characterize the polarimeter (\autoref{sec:initial-results})
    \item Used the results of the initial characterization to design, build, and calibrate a dual rotating quarter-wave plate compensator (\autoref{sec:dqwp})
    \item Characterized the polarimeter performance with the DQWP, showing significant improvements in polarimetric efficiency (+17.5\% on average) and instrumental polarization (-5.4\% on average) (\autoref{sec:post-dqwp-characterization})
    \item Demonstrated the efficacy of the polarimeter with on-sky imaging of the circumstellar debris disk around HR 4796, showing one of the clearest views of the near, forward-scattering side of the disk (\autoref{sec:on-sky-results})
\end{itemize}

Future work includes the deployment of a new PBS which should resolve the polarimetric efficiency issues remaining in the r' filter. We will, once again, characterize the instrumental response following those changes, and we will recalibrate the DQWP. We also plan to utilize on-sky measurements of polarized and unpolarized standard stars to fit a model for the telescope M3. Finally, we plan to image a nearby bright-star in polarimetry to measure the polarimetric contrast curve, determining how close to the photon-noise limit the polarimeter can perform at different separations from the stellar PSF.

This work, especially the DQWP compensator, helps pave a path for future polarimeters on extremely large telescopes like GMT (\citenum{johns_giant_2012}) and ELT (\citenum{gilmozzi_european_2007}). The optical complexity of the high-contrast instruments on those telescopes are even less than ideal compared to current-generation instruments, which makes polarimetry even more difficult. Furthermore, the high angular resolutions produced by such giant aperture diameters will manifest new problems, such as polarimetric aberrations (\citenum{van_holstein_polarization-dependent_2023}), which will need to be removed for effective diffraction control to enable imaging of reflected-light exoplanets (regardless of the inclusion of a polarimeter). Therefore, the lessons learned on the current generation of instruments will be key for building the next-generation.

\acknowledgments 

M.L. is supported by the Alan Brass Prize Fellowship in Instrumentation and Technology Development at Steward Observatory. We are thankful to the director of Steward Observatory for discretionary funding to support this work. We made use of MagAO-X, which is grateful for support from the NSF MRI Award \#1625441. The Phase II upgrade program is made possible by the generous support of the Heising-Simons Foundation. MagAO-X uses the CACAO software package, which is supported by NSF Award \#2410616. This paper includes data gathered with the 6.5 meter Magellan Telescopes located at Las Campanas Observatory, Chile. We are grateful for the support of the technical staff and the telescope operators who made this work possible. J.N.A was supported by NASA through the NASA Hubble Fellowship grant \#HST-HF2-51547.001-A awarded by the Space Telescope Science Institute, which is operated by the Association of Universities for Research in Astronomy. B.L.L. acknowledges support from the National Science Foundation Astronomy \& Astrophysics Postdoctoral Fellowship under Award No. 2401654. Any opinions, findings, and conclusions or recommendations expressed in this material are those of the author(s) and do not necessarily reflect the views of the National Science Foundation.

\bibliography{report} 

\begin{thebibliography}{10}

\bibitem{sparks_imaging_2002}
Sparks, W.~B. and Ford, H.~C., ``Imaging {Spectroscopy} for {Extrasolar}
  {Planet} {Detection},'' {\em Astrophys. J.}~{\bf 578},  543 (Oct. 2002).

\bibitem{traub_direct_2010}
Traub, W.~A. and Oppenheimer, B.~R., ``Direct imaging of exoplanets,'' {\em
  Exoplanets} ,  111--156 (Dec. 2010).

\bibitem{bowler_imaging_2016}
Bowler, B.~P., ``Imaging extrasolar giant planets,'' {\em Publ. Astron. Soc.
  Pac.}~{\bf 128},  102001 (Oct. 2016).
\newblock arXiv: 1605.02731.

\bibitem{benisty_optical_2023}
Benisty, M., Dominik, C., Follette, K., Garufi, A., Ginski, C., Hashimoto, J.,
  Keppler, M., Kley, W., and Monnier, J., ``Optical and near-infrared view of
  planet-forming disks and protoplanets,'' in [{\em Protostars and {Planets}
  {VII}}{\nolinebreak\hspace{0.1em}]},   {\bf 534},  605, Protostars and
  Planets VII (July 2023).
\newblock Conference Name: Protostars and Planets VII Place: eprint:
  arXiv:2203.09991 ADS Bibcode: 2023ASPC..534..605B.

\bibitem{andrews_observations_2020}
Andrews, S.~M., ``Observations of {Protoplanetary} {Disk} {Structures},'' {\em
  Annu. Rev. Astron. Astrophys.}~{\bf 58},  483--528 (Aug. 2020).

\bibitem{kley_planet-disk_2012}
Kley, W. and Nelson, R.~P., ``Planet-{Disk} {Interaction} and {Orbital}
  {Evolution},'' {\em Annu. Rev. Astron. Astrophys.}~{\bf 50},  211--249 (Sept.
  2012).

\bibitem{dong_how_2017}
Dong, R. and Fung, J., ``How {Bright} are {Planet}-induced {Spiral} {Arms} in
  {Scattered} {Light}?,'' {\em Astrophys. J.}~{\bf 835},  38 (Jan. 2017).
\newblock ADS Bibcode: 2017ApJ...835...38D.

\bibitem{dong_what_2017}
Dong, R. and Fung, J., ``What is the {Mass} of a {Gap}-opening {Planet}?,''
  {\em Astrophys. J.}~{\bf 835},  146 (Feb. 2017).
\newblock ADS Bibcode: 2017ApJ...835..146D.

\bibitem{guyon_extreme_2018}
Guyon, O., ``Extreme adaptive optics,'' {\em Annu. Rev. Astron.
  Astrophys.}~{\bf 56},  315--355 (Sept. 2018).

\bibitem{tazaki_effect_2019}
Tazaki, R., Tanaka, H., Muto, T., Kataoka, A., and Okuzumi, S., ``Effect of
  dust size and structure on scattered-light images of protoplanetary discs,''
  {\em MNRAS}~{\bf 485},  4951--4966 (June 2019).
\newblock ADS Bibcode: 2019MNRAS.485.4951T.

\bibitem{kuhn_imaging_2001}
Kuhn, J.~R., Potter, D., and Parise, B., ``Imaging polarimetric observations of
  a new circumstellar disk system,'' {\em Astrophys. J.}~{\bf 553},  L189 (May
  2001).

\bibitem{lafreniere_new_2007}
Lafreni{\`e}re, D., Marois, C., Doyon, R., Nadeau, D., and Artigau, {\'E}., ``A
  new algorithm for point-spread function subtraction in high-contrast imaging:
  a demonstration with angular differential imaging,'' {\em Astrophys. J.}~{\bf
  660},  770--780 (May 2007).

\bibitem{soummer_detection_2012}
Soummer, R., Pueyo, L., and Larkin, J., ``Detection and characterization of
  exoplanets and disks using projections on karhunen-lo{\`e}ve eigenimages,''
  {\em Astrophys. J. Lett.}~{\bf 755},  L28 (Aug. 2012).

\bibitem{monnier_multiple_2019}
Monnier, J.~D., Harries, T.~J., Bae, J., Setterholm, B.~R., Laws, A., Aarnio,
  A., Adams, F.~C., Andrews, S., Calvet, N., Espaillat, C., Hartmann, L.,
  Kraus, S., McClure, M., Miller, C., Oppenheimer, R., Wilner, D., and Zhu, Z.,
  ``Multiple spiral arms in the disk around intermediate-mass binary {HD}
  {34700A},'' {\em Astrophys. J.}~{\bf 872},  122 (Feb. 2019).

\bibitem{tschudi_quantitative_2021}
Tschudi, C. and Schmid, H.~M., ``Quantitative polarimetry of the disk around
  {HD} 169142,'' {\em A \& A}~{\bf 655},  A37 (Nov. 2021).

\bibitem{ma_quantitative_2023}
Ma, J., Schmid, H.~M., and Tschudi, C., ``Quantitative polarimetry for the
  transition disk in {RX} {J1604}.3-213010,'' {\em A \& A}~{\bf 676},  A6 (Aug.
  2023).

\bibitem{tinbergen_accurate_2007}
Tinbergen, J., ``Accurate optical polarimetry on the nasmyth platform,'' {\em
  Publ. Astron. Soc. Pac.}~{\bf 119}(862),  1371--1384 (2007).

\bibitem{perrin_polarimetry_2015}
Perrin, M.~D., Duchene, G., Millar-Blanchaer, M., Fitzgerald, M.~P., Graham,
  J.~R., Wiktorowicz, S.~J., Kalas, P.~G., Macintosh, B., Bauman, B., Cardwell,
  A., Chilcote, J., Rosa, R. J.~D., Dillon, D., Doyon, R., Dunn, J., Erikson,
  D., Gavel, D., Goodsell, S., Hartung, M., Hibon, P., Ingraham, P., Kerley,
  D., Konapacky, Q., Larkin, J.~E., Maire, J., Marchis, F., Marois, C., Mittal,
  T., Morzinski, K.~M., Oppenheimer, B.~R., Palmer, D.~W., Patience, J.,
  Poyneer, L., Pueyo, L., Rantakyr{\"o}, F.~T., Sadakuni, N., Saddlemyer, L.,
  Savransky, D., Soummer, R., Sivaramakrishnan, A., Song, I., Thomas, S.,
  Wallace, J.~K., Wang, J.~J., and Wolff, S.~G., ``Polarimetry with the gemini
  planet imager: methods, performance at first light, and the circumstellar
  ring around hr 4796a,'' {\em Astrophys. J.}~{\bf 799},  182 (Jan. 2015).

\bibitem{holstein_polarimetric_2020}
Holstein, R. G.~v., Girard, J.~H., Boer, J.~d., Snik, F., Milli, J., Stam,
  D.~M., Ginski, C., Mouillet, D., Wahhaj, Z., Schmid, H.~M., Keller, C.~U.,
  Langlois, M., Dohlen, K., Vigan, A., Pohl, A., Carbillet, M., Fantinel, D.,
  Maurel, D., Orign{\'e}, A., Petit, C., Ramos, J., Rigal, F., Sevin, A.,
  Boccaletti, A., Coroller, H.~L., Dominik, C., Henning, T., Lagadec, E.,
  M{\'e}nard, F., Turatto, M., Udry, S., Chauvin, G., Feldt, M., and Beuzit,
  J.-L., ``Polarimetric imaging mode of {VLT}/{SPHERE}/{IRDIS} - {II}.
  {Characterization} and correction of instrumental polarization effects,,''
  {\em A \& A}~{\bf 633},  A64 (Jan. 2020).

\bibitem{holstein_calibration_2020}
Holstein, R. G.~v., Bos, S.~P., Ruigrok, J., Lozi, J., Guyon, O., Norris, B.,
  Snik, F., Chilcote, J., Currie, T., Groff, T.~D., Hart, J.~t., Jovanovic, N.,
  Kasdin, J., Kudo, T., Martinache, F., Mazin, B., Sahoo, A., Tamura, M.,
  Vievard, S., Walter, A., and Zhang, J., ``Calibration of the instrumental
  polarization effects of {SCExAO}-{CHARIS}{\textquoteright}
  spectropolarimetric mode,'' in [{\em Ground-based and {Airborne}
  {Instrumentation} for {Astronomy} {VIII}}{\nolinebreak\hspace{0.1em}]},
  {\bf 11447},  1113--1126, SPIE (Dec. 2020).

\bibitem{zhang_characterizing_2023}
Zhang, M., Millar-Blanchaer, M., Safonov, B., Lucas, M., Lilley, L., Ashcraft,
  J., Norris, B., Lozi, J., Guyon, O., and Bottom, M., ``Characterizing the
  instrument polarization of {SCExAO} {VAMPIRES},'' in [{\em Techniques and
  {Instrumentation} for {Detection} of {Exoplanets}
  {XI}}{\nolinebreak\hspace{0.1em}]},   {\bf 12680},  261--282, SPIE (Oct.
  2023).

\bibitem{males_magao-x_2020}
Males, J.~R., Close, L.~M., Guyon, O., Hedglen, A.~D., Gorkom, K.~V., Long,
  J.~D., Kautz, M., Lumbres, J., Schatz, L., Rodack, A., Miller, K., Doelman,
  D., Snik, F., Bos, S., Knight, J.~M., Morzinski, K., Gasho, V., Keller, C.,
  Haffert, S., and Pearce, L., ``{MagAO}-{X} first light,'' in [{\em Adaptive
  {Optics} {Systems} {VII}}{\nolinebreak\hspace{0.1em}]},   {\bf 11448},
  918--925, SPIE (Dec. 2020).

\bibitem{souza_concept_2025}
Souza, T. G. B.~d., Close, L.~M., Li, J., Pereira, R., Haffert, S.~Y., Males,
  J.~R., Kueny, J., Kautz, M.~Y., Long, J.~D., Liberman, J., Twitchell, K.,
  Johnson, P., Guyon, O., McEwen, E., Hedglen, A.~D., Tonucci, E., and Mars,
  M., ``Concept, implementation, and on-sky commissioning of the {AO}
  polarimetric module on {MagAO}-{X},'' in [{\em Techniques and
  {Instrumentation} for {Detection} of {Exoplanets}
  {XII}}{\nolinebreak\hspace{0.1em}]},   {\bf 13627},  748--756, SPIE (Sept.
  2025).

\bibitem{mcintosh_characterizing_2026}
McIntosh, T., Zhang, M., Lewis, B.~L., Lucas, M., and Millar-Blanchaer, M.~A.,
  ``Updating the scexao/charis polarimetric calibration following the nasmyth
  beam-switcher upgrade,'' in [{\em Space {Telescopes} and {Instrumentation}
  2026: Adaptive Optics Systems}{\nolinebreak\hspace{0.1em}]},  SPIE (July
  2026).
\newblock arXiv:2607.23616 [astro-ph.IM].

\bibitem{norris_vampires_2015}
Norris, B., Schworer, G., Tuthill, P., Jovanovic, N., Guyon, O., Stewart, P.,
  and Martinache, F., ``The {VAMPIRES} instrument: imaging the innermost
  regions of protoplanetary discs with polarimetric interferometry,'' {\em
  MNRAS}~{\bf 447},  2894--2906 (Mar. 2015).

\bibitem{joost_t_hart_full_2021}
Joost~`t Hart, G.~J., van Holstein, R.~G., Bos, S.~P., Ruigrok, J., Snik, F.,
  Lozi, J., Guyon, O., Kudo, T., Zhang, J., Jovanovic, N., Norris, B.,
  Martinod, M.-A., Groff, T.~D., Chilcote, J., Currie, T., Tamura, M., Vievard,
  S., Sahoo, A., Deo, V., Ahn, K., Martinache, F., and Kasdin, J., ``Full
  characterization of the instrumental polarization effects of the
  spectropolarimetric mode of {SCExAO}-{CHARIS},'' (Aug. 2021).
\newblock Publication Title: arXiv e-prints ADS Bibcode: 2021arXiv210804833J.

\bibitem{lucas_visible-light_2024}
Lucas, M., Norris, B., Guyon, O., Bottom, M., Deo, V., Vievard, S., Lozi, J.,
  Ahn, K., Ashcraft, J., Currie, T., Doelman, D., Kudo, T., Leboulleux, L.,
  Lilley, L., Millar-Blanchaer, M., Safonov, B., Tuthill, P., Uyama, T., Walk,
  A., and Zhang, M., ``Visible-light high-contrast imaging and polarimetry with
  {SCExAO}/{VAMPIRES},'' {\em Publ. Astron. Soc. Pac.}~{\bf 136},  114504 (Nov.
  2024).

\bibitem{hinkley_speckle_2009}
Hinkley, S., Oppenheimer, B.~R., Soummer, R., Brenner, D., Graham, J.~R.,
  Perrin, M.~D., Sivaramakrishnan, A., Lloyd, J.~P., Roberts, L.~C., and Kuhn,
  J., ``Speckle suppression through dual imaging polarimetry, and a
  ground-based image of the hr 4796a circumstellar disk,'' {\em Astrophys.
  J.}~{\bf 701},  804 (July 2009).

\bibitem{milli_near-infrared_2017}
Milli, J., Vigan, A., Mouillet, D., Lagrange, A.-M., Augereau, J.-C., Pinte,
  C., Mawet, D., Schmid, H.~M., Boccaletti, A., Matr{\`a}, L., Kral, Q., Ertel,
  S., Chauvin, G., Bazzon, A., M{\'e}nard, F., Beuzit, J.-L., Thalmann, C.,
  Dominik, C., Feldt, M., Henning, T., Min, M., Girard, J.~H., Galicher, R.,
  Bonnefoy, M., Fusco, T., Boer, J.~d., Janson, M., Maire, A.-L., Mesa, D., and
  Schlieder, J.~E., ``Near-infrared scattered light properties of the {HR} 4796
  {A} dust ring - {A} measured scattering phase function from 13.6{\textdegree}
  to 166.6{\textdegree},'' {\em Astron. Astrophys.}~{\bf 599},  A108 (Mar.
  2017).

\bibitem{milli_optical_2019}
Milli, J., Engler, N., Schmid, H.~M., Olofsson, J., M{\'e}nard, F., Kral, Q.,
  Boccaletti, A., Th{\'e}bault, P., Choquet, E., Mouillet, D., Lagrange, A.-M.,
  Augereau, J.-C., Pinte, C., Chauvin, G., Dominik, C., Perrot, C., Zurlo, A.,
  Henning, T., Beuzit, J.-L., Avenhaus, H., Bazzon, A., Moulin, T., Llored, M.,
  Moeller-Nilsson, O., Roelfsema, R., and Pragt, J., ``Optical polarised phase
  function of the {HR} {4796A} dust ring,'' {\em A \& A}~{\bf 626},  A54 (June
  2019).

\bibitem{sahoo_precision_2020}
Sahoo, A., Guyon, O., Lozi, J., Chilcote, J., Jovanovic, N., Brandt, T., Groff,
  T., and Martinache, F., ``Precision photometric and astrometric calibration
  using alternating satellite speckles,'' {\em Astron. J.}~{\bf 159},  250 (May
  2020).

\bibitem{twitchell_dynamic_2026}
Twitchell, K., Haffert, S., Males, J., Close, L., and Lucas, M., ``Closed-loop
  atmospheric dispersion correction for high contrast imaging with magao-x,''
  in [{\em Space {Telescopes} and {Instrumentation} 2026: Adaptive Optics
  Systems}{\nolinebreak\hspace{0.1em}]},  SPIE (in prep 2026).

\bibitem{savransky_computer_2013}
Savransky, D., Thomas, S.~J., Poyneer, L.~A., and Macintosh, B.~A., ``Computer
  vision applications for coronagraphic optical alignment and image
  processing,'' {\em Appl. Opt.}~{\bf 52},  3394 (May 2013).
\newblock ADS Bibcode: 2013ApOpt..52.3394S.

\bibitem{follette_seeds_2015}
Follette, K.~B., Grady, C.~A., Swearingen, J.~R., Sitko, M.~L., Champney,
  E.~H., van~der Marel, N., Takami, M., Kuchner, M.~J., Close, L.~M., Muto, T.,
  Mayama, S., McElwain, M.~W., Fukagawa, M., Maaskant, K., Min, M., Russell,
  R.~W., Kudo, T., Kusakabe, N., Hashimoto, J., Abe, L., Akiyama, E., Brandner,
  W., Brandt, T.~D., Carson, J., Currie, T., Egner, S.~E., Feldt, M., Goto, M.,
  Guyon, O., Hayano, Y., Hayashi, M., Hayashi, S., Henning, T., Hodapp, K.,
  Ishii, M., Iye, M., Janson, M., Kandori, R., Knapp, G.~R., Kuzuhara, M.,
  Kwon, J., Matsuo, T., Miyama, S., Morino, J.-I., Moro-Martin, A., Nishimura,
  T., Pyo, T.-S., Serabyn, E., Suenaga, T., Suto, H., Suzuki, R., Takahashi,
  Y., Takato, N., Terada, H., Thalmann, C., Tomono, D., Turner, E.~L.,
  Watanabe, M., Wisniewski, J.~P., Yamada, T., Takami, H., Usuda, T., and
  Tamura, M., ``{SEEDS} {ADAPTIVE} {OPTICS} {IMAGING} {OF} {THE} {ASYMMETRIC}
  {TRANSITION} {DISK} {OPH} {IRS} 48 {IN} {SCATTERED} {LIGHT}*,'' {\em
  Astrophys. J.}~{\bf 798},  132 (Jan. 2015).

\bibitem{avenhaus_disks_2018}
Avenhaus, H., Quanz, S.~P., Garufi, A., Perez, S., Casassus, S., Pinte, C.,
  Bertrang, G. H.~M., Caceres, C., Benisty, M., and Dominik, C., ``Disks around
  {T} tauri stars with {SPHERE} ({DARTTS}-{S}). {I}. {SPHERE}/{IRDIS}
  polarimetric imaging of eight prominent {T} tauri disks,'' {\em Astrophys.
  J.}~{\bf 863},  44 (Aug. 2018).
\newblock ADS Bibcode: 2018ApJ...863...44A.

\bibitem{de_regt_polarimetric_2024}
de~Regt, S., Ginski, C., Kenworthy, M.~A., Caceres, C., Garufi, A., Gledhill,
  T.~M., Hales, A.~S., Huelamo, N., K{\'o}sp{\'a}l, {\'A}., Millar-Blanchaer,
  M.~A., P{\'e}rez, S., and Schreiber, M.~R., ``Polarimetric differential
  imaging with {VLT}/{NACO} - {A} comprehensive {PDI} pipeline for {NACO} data
  ({PIPPIN}),'' {\em Astron. Astrophys.}~{\bf 684},  A73 (Apr. 2024).

\bibitem{schmid_limb_2006}
Schmid, H.~M., Joos, F., and Tschan, D., ``Limb polarization of {Uranus} and
  neptune - {I}. {Imaging} polarimetry and comparison with analytic models,''
  {\em A \& A}~{\bf 452},  657--668 (June 2006).
\newblock Number: 2.

\bibitem{johns_giant_2012}
Johns, M., McCarthy, P., Raybould, K., Bouchez, A., Farahani, A., Filgueira,
  J., Jacoby, G., Shectman, S., and Sheehan, M., ``Giant {Magellan}
  {Telescope}: overview,'' in [{\em Ground-based and {Airborne} {Telescopes}
  {IV}}{\nolinebreak\hspace{0.1em}]},   {\bf 8444},  526--541, SPIE (Sept.
  2012).

\bibitem{gilmozzi_european_2007}
Gilmozzi, R. and Spyromilio, J., ``The {European} {Extremely} {Large}
  {Telescope} ({E}-{ELT}),'' {\em Messenger}~{\bf 127},  11 (Mar. 2007).
\newblock ADS Bibcode: 2007Msngr.127...11G.

\bibitem{van_holstein_polarization-dependent_2023}
van Holstein, R.~G., Keller, C.~U., Snik, F., and Bos, S.~P.,
  ``Polarization-dependent beam shifts upon metallic reflection in
  high-contrast imagers and telescopes,'' {\em Astronomy and Astrophysics}~{\bf
  677},  A150 (Sept. 2023).
\newblock ADS Bibcode: 2023A\&A...677A.150V.

\end{thebibliography}
\bibliographystyle{spiebib} 

\appendix
\section{Mueller Matrices}\label{sec:mueller-matrices}

Here we describe the Mueller matrices used in this work.

\subsection{Rotation matrix}
The rotation matrix is used to alter the reference frame.
\begin{equation}
    \mathbf{T}\left( \theta \right) =
    \begin{pmatrix}
    1 & 0 & 0 & 0 \\
    0 & \cos{2\theta} & \sin{2\theta} & 0 \\
    0 & -\sin{2\theta} & \cos{2\theta} & 0 \\
    0 & 0 & 0 & 1
    \end{pmatrix}
\end{equation}
For example, an arbitrary Mueller matrix oriented at angle $\theta$ is described by
\begin{equation}
    \mathbf{T}^{-1}\left(\theta\right) \cdot \mathbf{M} \cdot\mathbf{T}\left(\theta\right).
\end{equation}
Note that $\mathbf{T}^{-1}\left(\theta\right) = \mathbf{T}^T\left(\theta\right) = \mathbf{T}\left(-\theta\right)$.

\subsection{Linear diattenuator}
A component with linear diattenuation $\chi$
\begin{equation}
    \mathbf{M}_\text{diat}\left(\chi\right) =
    \begin{pmatrix}
    1 & \chi & 0 & 0 \\
    \chi & 1 & 0 & 0 \\
    0 & 0 & \sqrt{1 - \chi^2} & 0 \\
    0 & 0 & 0 & \sqrt{1 - \chi^2}
    \end{pmatrix}
\end{equation}

\subsection{Linear retarder}
A component with linear retardance $\eta$
\begin{equation}
    \mathbf{M}_\text{ret}\left(\eta\right) =
    \begin{pmatrix}
    1 & 0 & 0 & 0 \\
    0 & 1 & 0 & 0 \\
    0 & 0 & \cos\eta & \sin\eta \\
    0 & 0 & -\sin\eta & \cos\eta
    \end{pmatrix}
\end{equation}

\subsection{Elliptical retarder}
A component with different retardance for each eigenpolarization, $\eta^Q$, $\eta^U$, and $\eta^V$
\begin{equation}
    \mathbf{M}_\text{ret}\left(\eta^Q, \eta^U, \eta^V\right) =
    \begin{pmatrix}
    1 & 0 & 0 & 0 \\[2pt]
    0 & \alpha + (1-\alpha)\dfrac{(\eta^Q)^2}{\etamag^2}
    & (1-\alpha)\dfrac{\eta^Q\eta^U}{\etamag^2} - \beta\,\dfrac{\eta^V}{\etamag}
    & (1-\alpha)\dfrac{\eta^Q\eta^V}{\etamag^2} + \beta\,\dfrac{\eta^U}{\etamag} \\[8pt]
    0 & (1-\alpha)\dfrac{\eta^Q\eta^U}{\etamag^2} + \beta\,\dfrac{\eta^V}{\etamag}
    & \alpha + (1-\alpha)\dfrac{(\eta^U)^2}{\etamag^2}
    & (1-\alpha)\dfrac{\eta^U\eta^V}{\etamag^2} - \beta\,\dfrac{\eta^Q}{\etamag} \\[8pt]
    0 & (1-\alpha)\dfrac{\eta^Q\eta^V}{\etamag^2} - \beta\,\dfrac{\eta^U}{\etamag}
    & (1-\alpha)\dfrac{\eta^U\eta^V}{\etamag^2} + \beta\,\dfrac{\eta^Q}{\etamag}
    & \alpha + (1-\alpha)\dfrac{(\eta^V)^2}{\etamag^2}
    \end{pmatrix}
\end{equation}
where
\begin{equation}
    \etamag = \sqrt{\left(\eta^Q\right)^2 + \left(\eta^U\right)^2 + \left(\eta^V\right)^2},
    \quad \alpha = \cos\etamag, \quad \beta = \sin\etamag.
\end{equation}

\subsection{MagAO-X Mueller Matrices}\label{sec:magaox-mueller-matrices}
For each MagAO-X optical term we use the following combinations of Mueller matrices:
\begin{equation}
    \mathbf{M}_\text{Tel}\left(a, p \left| \chi_\text{M3}, \eta_\text{M3}\right.\right) = 
    \mathbf{T}\left(a\right)\cdot
    \mathbf{M}_\text{diat}\left(\chi_\text{M3}\right)\cdot
    \mathbf{M}_\text{ret}\left(\eta_\text{M3}\right)\cdot
    \mathbf{T}\left(p\right)
\end{equation}

\begin{equation}
    \mathbf{M}_\text{HWP}\left(\theta_\text{HWP} \left| \delta_\text{HWP}, \eta_\text{HWP}\right.\right) = 
    \mathbf{T}^{-1}\left(\theta_\text{HWP} + \delta_\text{HWP}\right)\cdot
    \mathbf{M}_\text{ret}\left(\eta_\text{HWP}\right)\cdot
    \mathbf{T}\left(\theta_\text{HWP} + \delta_\text{HWP}\right)
\end{equation}

\begin{equation}
    \mathbf{M}_\text{M4}\left(\left|\chi_\text{M4}, \eta_\text{M4}\right.\right) = \mathbf{M}_\text{diat}\left(\chi_\text{M4}\right)\cdot\mathbf{M}_\text{ret}\left(\eta_\text{M4}\right)
\end{equation}

\begin{equation}
    \begin{aligned}
    \mathbf{M}_\text{IMR}\left(\theta_\text{IMR} \left| \delta_\text{IMR}, \chi_\text{IMR}, \eta^Q_\text{IMR}, \eta^U_\text{IMR}, \eta^V_\text{IMR}\right.\right) & = \\
    \mathbf{T}^{-1}\left(\theta_\text{IMR} + \delta_\text{IMR}\right)\cdot 
    \mathbf{M}_\text{diat}\left(\chi_\text{IMR}\right)\cdot &
    \mathbf{M}_\text{ret}\left(\eta^Q_\text{IMR}, \eta^U_\text{IMR}, \eta^V_\text{IMR}\right)\cdot
    \mathbf{T}\left(\theta_\text{IMR} + \delta_\text{IMR}\right)
    \end{aligned}
\end{equation}

\begin{equation}
    \mathbf{M}_\text{Inst}\left(\left|\theta_\text{Inst}, \chi_\text{Inst}, \eta^Q_\text{Inst}, \eta^U_\text{Inst}, \eta^V_\text{Inst}\right.\right) = 
    \mathbf{T}^{-1}\left(\theta_\text{Inst}\right)\cdot
    \mathbf{M}_\text{diat}\left(\chi_\text{Inst}\right)\cdot
    \mathbf{M}_\text{ret}\left(\eta^Q_\text{Inst}, \eta^U_\text{Inst}, \eta^V_\text{Inst}\right)\cdot
    \mathbf{T}\left(\theta_\text{Inst}\right)
\end{equation}

\begin{equation}
    \mathbf{M}_\text{PBS}\left(\left|\chi_1, \chi_2\right.\right) = 
    \begin{cases}
        \mathbf{M}_\text{diat}\left(\chi_{1}\right) &\text{if \texttt{camsci1}} \\
        \mathbf{M}_\text{diat}\left(\chi_{2}\right) &\text{if \texttt{camsci2}}
    \end{cases}
\end{equation}

\end{document}